\documentclass[prd,nofootinbib,showpacs,superscriptaddress]{revtex4}
\pdfoutput=1

\usepackage{amsfonts}
\usepackage{amsmath}
\usepackage{amssymb}

\usepackage{graphicx}
\usepackage{dcolumn}
\usepackage{epsfig}
\usepackage[FIGTOPCAP]{subfigure}
\usepackage{epstopdf}
\usepackage{dcolumn}
\usepackage{bm}
\usepackage[latin1]{inputenc}
\usepackage{latexsym}
\usepackage{rotating}
\usepackage{hyperref}
\hypersetup{colorlinks=true,linkcolor=blue,citecolor=blue}
\usepackage[usenames]{color}
\usepackage{float}
\usepackage{xspace} 
\usepackage{mathrsfs}
\usepackage{enumitem}
\usepackage{caption}
\usepackage{ulem}
\def\fun#1#2{\lower3.6pt\vbox{\baselineskip0pt\lineskip.9pt
\ialign{$\mathsurround=0pt#1\hfil##\hfil$\crcr#2\crcr\sim\crcr}}}

\def\kms{\mathrm{km\,s}^{-1}}

\def\mass{{\cal M}}

\def\msun{{\mass_\odot}}

\def\beq{\begin{equation}}
\def\eeq{\end{equation}}
\def\barr{\begin{eqnarray}}
\def\earr{\end{eqnarray}}
\def\bsub{\begin{subequations}}
\def\esub{\end{subequations}}

\def\mgal{M_{\rm gal}}
\def\sbh{SBH}

\def\sbhs{SBHs}

\def\tcoal{t_{\rm coal}}

\def\ayr{A_{\rm yr}}

\def\zmax{z_{\rm max}}
\def\mmax{M_{\rm 12,max}}
\def\mmin{M_{\rm 12,min}}
\def\qmin{q_{\rm min}}

\usepackage{setspace}

\begin{document}

\title{Evolution Of Massive Black Hole Binaries In Rotating Stellar Nuclei: Implications For Gravitational Wave Detection}

\author{Alexander Rasskazov and David Merritt}
\affiliation{School of Physics and Astronomy and Center for Computational Relativity and Gravitation, Rochester Institute of Technology, Rochester, NY 14623}

\date{\today}

\begin{abstract}
We compute the isotropic gravitational wave (GW) background produced by binary supermassive
black holes (\sbhs) in galactic nuclei. In our model, massive binaries evolve at early times
via gravitational-slingshot interaction with nearby stars, and at later times by the emission of GWs.
Our expressions for the rate of binary hardening in the ``stellar" regime are taken from the recent
work of Vasiliev et al., who show that in the non-axisymmetric galaxies expected to form via mergers,
stars are supplied to the center at high enough rates to ensure binary coalescence on Gyr timescales.
We also include, for the first time, the extra degrees of freedom associated with evolution of the binary's
orbital plane;
in rotating nuclei, interaction with stars causes the orientation and the eccentricity of a massive binary
to change in tandem, leading in some cases to very high eccentricities ($e>0.9$) before the binary
enters the GW-dominated regime.
We argue that previous studies have over-estimated the mean ratio of \sbh\ mass to galaxy bulge mass
by factors of 2 -- 3.
In the frequency regime currently accessible to pulsar timing arrays (PTAs), our assumptions imply a
factor 2 -- 3 reduction in the characteristic strain compared with the values computed in most recent studies,
removing the tension that currently exists between model predictions and the non-detection of GWs.
\end{abstract}

\pacs{04.30.-w, 04.30.Tv, 97.60.Lf}
\maketitle
\allowdisplaybreaks[4]


\section{Introduction}
\label{Section:Intro}

Pulsar timing arrays (PTAs \cite{Foster1990}) are designed to detect the low-frequency ($\sim$nHz) gravitational wave (GW) background
that would be generated by a population of binary supermassive black holes (SBHs).
In the simplest model -- a cosmologically homogeneous and isotropic population of massive binaries on circular orbits, 
which evolve solely due to GW emission --
the characteristic strain of the GW-induced distortions has a frequency dependence given by \citep{Phinney2001}
\begin{equation}
h_c(f) = \ayr \left(\frac{f}{f_\mathrm{yr}}\right)^{-2/3}
\label{Equation:hcPhinney}
\end{equation}
where $f_\mathrm{yr}\equiv 1/\mathrm{yr}$.
The corresponding energy density per unit logarithmic frequency, expressed in terms of the cosmological critical density 
$\rho_c= 3H_0^2/(8\pi G)$, is \citep{Phinney2001}
\begin{equation}
\Omega_\mathrm{GW}(f) \equiv \frac{1}{\rho_c} \frac{d\rho_\mathrm{GW}}{d \ln f} = \frac{2\pi^2}{3H_0^2} f^2 h_c^2(f)
\label{Equation:OmegaGW}
\end{equation}
where $H_0\approx 7.2\times 10^{-11}$ yr$^{-1}$ is the Hubble constant.
The parameter $A_\mathrm{yr}$, the predicted strain at a frequency of one inverse year, depends in a possibly complicated
way on the astrophysical
parameters that characterize the binary population, including the mass function of \sbhs; the distribution of binary mass
ratios; the galaxy merger rate; and the rate at which binaries attain separations small enough ($\lesssim 10^{-2}$ pc)
that GW emission can dominate their evolution.
Theoretical estimates of $A_\mathrm{yr}$ typically lie in the range $\sim 10^{-15} - 10^{-14}$ 
\citep{Sesana2013a,McWilliams2014,Ravi2014}.
Detection of GWs in this frequency regime would provide robust evidence for the existence of binary \sbhs\
and would allow the astrophysical parameters
that determine the frequency spectrum to be constrained \cite{Sesana2013b}.

The peak sensitivity of a PTA occurs at a frequency that is roughly the inverse of the time over which pulsar timing data
has been collected \citep{Cordes2013}.
That time is now roughly one decade.
At these lower frequencies, $f\ll f_\mathrm{yr}$,
the characteristic strain is expected to differ from the prediction of Eq. (\ref{Equation:hcPhinney}).
The semimajor axis of a binary \sbh\ with orbital period $P$ is 
\begin{eqnarray}
a = 1.0\times 10^{-2} \left(\frac{M_{12}}{10^8 \msun}\right)^{1/3} \left(\frac{P}{10\;\mathrm{yr}}\right)^{2/3}
\mathrm{pc} 
\end{eqnarray}
with $M_{12}=M_1+M_2$ the mass of the binary.
At separations corresponding to orbital frequencies less than $\sim (10\;\mathrm{yr})^{-1}$, i.e. $a\gtrsim 10^{-2}$ pc,
a massive binary is expected to evolve primarily via interactions with ambient stars and gas in the galactic nucleus
rather than by GW emission \citep[][chapter 8]{DEGN}.
Furthermore there is no compelling reason why binaries at these large separations should be on circular orbits; it is only
at smaller separations that GW emission becomes effective at reducing eccentricities.
Both considerations would predict a reduction in $h_c$ below a certain frequency, compared with Eq.~(\ref{Equation:hcPhinney}).
\citet{Sampson2015} suggested a simple parametrization of the GW spectrum describing a population of 
binaries that have been ``environmentally'' influenced:
\begin{subequations}
\label{Equation:Sampsonhc}
\begin{eqnarray}
h_c(f) &=& A \frac{\left(f/f_\mathrm{yr}\right)^{-2/3}}{\left[1+\left(f_\mathrm{bend}/f\right)^\kappa\right]^{1/2}}, \\
A &=& \ayr \left[1+\left(f_\mathrm{bend}/f_\mathrm{yr}\right)^\kappa\right]^{1/2}
\end{eqnarray}
\end{subequations}
where $f_\mathrm{bend}$ is understood as the  orbital frequency below which environmental interactions
dominate the binary's evolution and $\kappa$ is determined by the type of interaction; 
in the case of stellar scattering considered in this paper, $\kappa=10/3$.
Simple evolutionary models suggest $f_\mathrm{bend}\approx 10^{-9}$ Hz \citep{Sesana2013b}, a frequency regime
that is beginning to be probed by PTAs.

Analysis of pulsar timing data by three groups
has so far succeeded in placing only upper limits on $\ayr$:
$3.0\times 10^{-15}$ (EPTA \cite{Lentati2015}); $1.0\times 10^{-15}$ (PPTA \cite{Shannon2015}); and
$1.5\times 10^{-15}$ (NANOGrav \cite{Arzoumanian2016}).
These values are generally interpreted as being ``in tension with'' the predictions of some theoretical models; for instance, \citet{McWilliams2014} predict $\ayr\approx 10^{-14.5}$. According to \citet[Table S8]{Shannon2015}, even the models with the lowest predicted $\ayr$ (\cite{Sesana2013a} and \cite{Ravi2014}) have only 9\% and 21\% probability, respectively, of being consistent with the limit derived from their observations.

In the present paper we present a new calculation of $h_c(f)$.
Our treatment differs in three important ways from previous ones.

\medskip

1. {\it Binary hardening rates.} The ``final-parsec problem'' \citep{MM2003} refers to the possibility that massive
binaries might stall at separations much greater than required for the emission of detectable GWs.
Here we make use of recent work \citep{Vasiliev2014,Vasiliev2015} 
which shows that even in ``collisionless'' (gas-free, long-relaxation-time) nuclei, like those of massive galaxies, a
modest departure from exact axisymmetry is sufficient to keep a massive binary shrinking.
In such a nucleus, the binary hardening rate decreases with time, but interactions with stars are nevertheless
able to drive the binary to coalescence on a timescale of order 1 Gyr or less. 
In an accurately axisymmetric nucleus, hardening rates are low enough that coalescence is not likely in a Hubble time,
but a binary can still enter into the PTA frequency regime.
Guided by the observations, we characterize high-luminosity bulges as triaxial and low-luminosity bulges as
axisymmetric, then use the expressions derived in the cited papers to compute binary hardening rates for the 
different galaxy populations.

\medskip

2. {\it Eccentricity evolution.} In the regime where binary hardening is driven by interaction with stars,
as opposed to GW emission, eccentricity evolution has been shown to be modest, 
at least in spherical nonrotating nuclei \citep{Quinlan1996}.
On this basis, most discussion of the stochastic GW spectrum have assumed zero eccentricities.
The situation can be very different in the case of nuclei with significant rotation, particularly if
the angular momentum of the massive binary is initially misaligned with that of the nucleus \citep{PaperI}.
Given such initial conditions, the binary's orbital plane rotates to bring its angular momentum vector more in
alignment with the nuclear rotation axis, and the binary's eccentricity simultaneously increases, sometimes
to very large values ($e>0.9$). 
As the binary aligns fully with the nucleus, its eccentricity returns again to lower values.
Our models are the first to include these additional degrees of freedom.

\medskip

3. {\it SBH demographics.} \citet{Arzoumanian2016} adopted prior probability distributions for
$\ayr$ from the modeling studies of \citet[][S13]{Sesana2013a} and \citet[][MOP]{McWilliams2014} and 
used them to infer posterior distributions of the parameters $A$, $f_\mathrm{bend}$ and $\kappa$ in Eq.~(\ref{Equation:Sampsonhc}). 
MOP assumed a mean ratio of \sbh\ mass
to bulge mass of $\sim 0.003$, and they further augmented the \sbh\ mass function to account for a putative population of
``overmassive'' \sbhs\ in giant galaxies.
The resulting estimate of $\ayr\approx 10^{-14.4}$ was found by Arzoumanian et al.
to be difficult to reconcile with the PTA data.
In order to limit the predicted contribution at low frequencies, a value $f_\mathrm{bend} \gtrsim 10^{-8}$ Hz was required,
substantially larger than the value expected physically unless nuclear densities in giant galaxies exceed
$\sim 10^{-3}\msun\mathrm{pc}^{-3}$. 
This possibility was judged unlikely by Arzoumanian et al., 
and those authors suggested that the MOP prior might be in error (too large),
either because binary \sbhs\ typically stall at separations outside the PTA band, 
or because the characterization adopted by MOP for the parent population of
\sbhs\ was somehow incorrect. \citet{Shannon2015} reached a similar conclusion.
We argue in fact that both S13 and MOP substantially overestimated the mean ratio of \sbh\ mass to bulge mass.
A more conservative (in the sense of being based on more compelling data) estimate of this ratio is $ 0.001$
which is the fiducial value we adopt here.

\medskip

The last of these assumptions is most important at setting the predicted amplitude of $h_c(f)$ at frequencies that lie in the current range
of PTA sensitivity.
Our models have $\ayr < 10^{-15}$, a factor of at least two lower than in most other recent calculations of $h_c(f)$
\cite{Sesana2013a,McWilliams2014,Ravi2014}.
Although we make no attempt to model the PTA data in the manner of \citet{Arzoumanian2016} or \citet{Shannon2015},
such a low value for $\ayr$ would presumably
(1) be consistent with a physically more plausible range of parameters $\{A, f_\mathrm{bend},\kappa\}$ in Eq. (\ref{Equation:Sampsonhc});
(2) remove the ``tension" between the non-detection of a stochastic GW signal by the various groups and the predictions of MOP; and
(3) unfortunately, imply that a PTA detection of the stochastic GW background from inspiralling \sbhs\ is not likely to occur in
the immediate future.


Our physical model for the formation and evolution of massive binaries is presented in \S\ref{Section:Method}; 
this section also includes the derivation of a formula for $h_c(f)$ that, for the first time, 
allows for any possible functional form of the time-dependence of the binary hardening rate.
\S\ref{Section:Results} presents estimates of the characteristic strain spectrum and its dependence on the parameters
that define the initial population of binaries and their host galaxies.
\S\ref{Section:Discussion} sums up and 
discusses the implications of our results for the detection of isotropic gravitational wave background via PTAs.

\section{Method}
\label{Section:Method}

We assume that shortly after two galaxies merge, the two \sbhs\ form a ``hard binary''\footnote{Defined as a binary that ejects passing
stars at typical velocities greater than the escape velocity from the nucleus \citep{Heggie1975}.} at the center of the merger product.  
The components of the binary have masses $M_1$ and $M_2$;
$M_{12} = M_1 + M_2$, $\mu_{12} = M_1M_2/\left(M_1+M_2\right)$ 
is the binary's reduced mass and $q\equiv M_2/M_1\leq 1$ its mass ratio (or $\mathcal{Q}= \mu/M_{12}=q/(1+q)^2$ its symmetric mass ratio). 
The initial semimajor axis is the ``hard-binary separation'' $a_h$:
\beq\label{Equation:a_h}
a_h \equiv \frac{G\mu}{4\sigma^2} \approx 2.7 \,
\mathcal{Q}\, \frac{M_{12}}{10^8 \msun} 
\left(\frac{\sigma}{200\,\kms}\right) ^{-2} \mathrm{pc}
\eeq
where $\sigma$ is the 1d stellar velocity dispersion in the nucleus,
and the initial orbital frequency is
\begin{subequations}\label{Equation:KeplerP}
\begin{eqnarray}
\frac{1}{P} &=& \frac{\sqrt{GM_{12}}}{2\pi a_h^{3/2}} = \frac{4\sigma^3}{\pi\mathcal{Q}^{3/2}GM_{12}} =\frac{4\sigma}{\pi\mathcal{Q}^{3/2}r_h} \\
&\approx&  0.8\times 10^{-12} \mathcal{Q}^{-3/2} \left(\frac{\sigma}{200\ \mathrm{km\ s}^{-1}}\right) \left(\frac{r_h}{10\ \mathrm{pc}}\right)^{-1} \mathrm{Hz}
\end{eqnarray}
\end{subequations}
where $r_h\equiv GM_{12}/\sigma^2$ is the binary's gravitational influence radius.
For reasonable values of the parameters in Eq.~(\ref{Equation:KeplerP}), 
this frequency is below the limit detectable by PTAs and so we ignore the contribution of binaries with $a>a_h$
to the GW background.
In what follows, we identify the galaxy merger rate with the rate of formation of binaries having $a=a_h$.

\subsection{Gravitational wave background from a population of massive binaries}
\label{Section:GW}

Consider the set of binaries that form, at any cosmological time, but with the same values of 
$\{M_1, M_2, \boldsymbol{L}\}$, where $L^2=GM_{12}a_h(1-e^2)$ and $e$ is the binary's
eccentricity at formation ($a=a_h$). After formation, the separation evolves as $a=a(t), t_h\le t \le t_c$ where $t_h$ is the formation time and
$t_c$ is the time at which the two \sbhs\ coalesce.
Define $N(a,t)da$ to be the number of binaries from this set, per unit comoving volume, 
that have separations in the interval $a$ to $a+da$ at time $t$ ($N$ can also depend on $M_1, M_2, \boldsymbol{L}$ and the properties of the host galaxy).
The function $N$ obeys a continuity equation:
\begin{eqnarray}
\frac{\partial N}{\partial t} + \frac{\partial}{\partial a} \left(N \overset{\bm .}{a}\right) &=& 0
\end{eqnarray}
with solution
\begin{eqnarray}\label{Equation:N}
N(a,t)\; da &=& \frac{\dot{\cal N}_m (t_h) }{\left|\overset{\bm .}{a}(a)\right |}\; da
\end{eqnarray}
with $\dot {\cal N}$ the galaxy merger rate:
\beq
\dot{\cal N}_m (t_h)\equiv -N(a_h,t_h)\, \overset{\bm .}{a}(a_h) 
\eeq
i.e. the rate, per comoving volume, at which galaxies are merging at time $t_h$ (we neglect the time it takes for a 
binary to become hard after the galaxy merger). 

By analogy with \citet{Ravi2014}, the specific intensity of GWs at the Earth from binaries with semimajor axes between $a$ and $a+da$ at redshifts between $z$ and $z+dz$ is
\begin{equation}
dI = \frac{L(f_{r},a)}{4\pi d_{L}^{2}}\frac{df_{r}}{df}N(a,z)\frac{dV_{c}}{dz}da\,dz,
\end{equation}
where $f$ is the observed GW frequency, $f_r=(1+z)f$ is the rest-frame frequency, $d_L$ is the luminosity distance,
\bsub\label{Equation:L}
\barr
L(f_r,a) &=& \frac{32}{5} \frac{G^{7/3}}{c^5} \mathcal{M}^{10/3} (2\pi f_{\rm orb})^{10/3} \displaystyle\sum_{n=1}^\infty g(n,e)\, \delta(f_r-nf_{\rm orb}),\\
\mathcal{M} &=& M_{12}\mathcal{Q}^{3/5},\\\
f_{\rm orb} &=& \frac{1}{2\pi}\sqrt\frac{GM_{12}}{a^3},\\
g(n,e) &=& \frac{n^4}{32} \Bigg[  \bigg\{J_{n-2}(ne) -2e J_{n-1}(ne) + \frac{2}{n} J_{n}(ne) + 2eJ_{n+1}(ne)-J_{n+2}(ne)\bigg\}^2\\\nonumber&+&\left(1-e^2\right)\bigg\{ J_{n-2}(ne) -2J_{n}(ne)+J_{n+2}(ne)\bigg\}^2 + \frac{4}{3n^2}J^2_{n}(ne)\Bigg]
\earr
\esub
is the GW luminosity per unit rest-frame frequency of a binary with semimajor axis $a$ and eccentricity $e$ 
(which is determined by $a$ since we are assuming here the same values for the other initial parameters) and 
\begin{equation}
\frac{dV_c}{dz} = \frac{4\pi cd_L^2}{H(z)(1+z)^2}
\end{equation}
is the comoving volume per unit $z$. Eq.~(\ref{Equation:L}) is taken from \cite{Peters1963}.

Now, the energy density in GWs at the Earth per logarithmic frequency unit is
\bsub
\barr
\Omega_{\rm GW}\rho_c c^2 &=& \frac{f}{c}\int_0^{\zmax}\int_{a_c}^{a_h}\frac{dI}{dadz}dadz \\
&=& f\int_0^{\zmax}\int_{a_c}^{a_h}\frac{L(f_r,a)N(a,z)}{H(z)(1+z)}dadz \\
&=& f\int_{t_r(\zmax)}^0 dt_r \int_{a_c}^{a_h} L(f_r,a) N(a,t_r)da\label{Equation:Omega_GW}
\earr
\esub
where $a_c$ is the orbital separation at which BHs coalesce and 
\bsub
\barr
t_r(z)&=&\int_z^0 \frac{dz'}{H(z')(1+z')},\\
H(z) &=& H_0 \left[\Omega_{\rm M}(1+z)^3+\Omega_{\Lambda}\right]^{1/2},\\
H_0 &=& 67.7 \,\mathrm{km/s/Mpc}, \ \ \Omega_{\rm M} = 0.31, \ \ \Omega_{\Lambda} = 0.69
\earr
\esub
is the proper time. 
Substituting the expressions (\ref{Equation:N}) for $N(a,t_r)$ and (\ref{Equation:L}) for $L(f_r,a)$ into 
Eq.~(\ref{Equation:Omega_GW}) and using the following property of the $\delta$ function:
\bsub
\barr
\int y(t)\delta(x(t)) dt &=& \frac{y(t_0)}{dx/dt(t_0)},\\
x(t_0) = 0,
\earr
\esub
we can carry out the integration over $a$, yielding
\barr
\Omega_{\rm GW}\rho_c c^2 &=& \frac{32}{5} \frac{G^{7/3}}{c^5} \mathcal{M}^{10/3} f\int_{t_r(\zmax)}^0 dt_r \, \displaystyle\sum_{a_c<a_n<a_h}\left(n\left|\frac{df_{\rm orb}}{da}(a_n)\right|\right)^{-1} \left(\frac{2\pi f_r}{n}\right)^{10/3} g(n,e(a_n)) \frac{\dot{\cal N}_m (t_h(a_n,t_r)) }{\left|\overset{\bm .}{a}(a_n) \right |}\nonumber \\
\earr
where $t_h(a,t_r)$ is the formation time of a binary that has semimajor axis $a$ at time $t_r$, and $a_n$ is the binary semimajor axis at which the orbital frequency is $f_r/n$, so that the $n$-th harmonic is contributing to the total energy radiated at frequency $f_r$:
\barr
f_{\rm orb} (a_n) = \frac{1}{2\pi}\sqrt\frac{GM_{12}}{a_n^3} = \frac{f_r}{n} = \frac{f(1+z(t_r))}{n}
\earr

It is straightforward to show that
\barr
\left(n\left|\frac{df_{\rm orb}}{da}(a_n)\right|\right)^{-1} = \frac{4\pi}{3n}\left(\frac{n}{2\pi f_r}\right)^{5/3}(GM_{12})^{1/3},
\earr
which gives us
\barr\label{Equation:OmegaGW0}
\Omega_{\rm GW}\rho_c c^2 &=& \frac{32}{5} \frac{G^{7/3}}{c^5} \mathcal{M}^{10/3} (GM_{12})^{1/3} f\int_{t_r(\zmax)}^0 dt_r \, \displaystyle\sum_{a_c<a_n<a_h}\frac{4\pi}{3n}\left(\frac{2\pi f_r}{n}\right)^{5/3} g(n,e(a_n)) \frac{\dot{\cal N}_m (t_h(a_n,t_r)) }{\left|\overset{\bm .}{a}(a_n)\right |}
\earr

We define the binary hardening rate, $S$, in the usual way as
\barr\label{Equation:DefineS}
S \equiv \frac{d}{dt}\left(\frac{1}{a}\right) 
\earr
and $S_h$ as the hardening rate in full loss-cone approximation (one that would occur if the distribution of stars in phase space were not affected by the presence of the binary; see Eq. \ref{Equation:S_h} and \ref{Equation:S_h(r_infl)}).
The initial hardening timescale (again, in full loss-cone approximation) is
\beq\label{Equation:t_h}
t_h \equiv \left|\frac{a}{\overset{\bm .}{a}}\right|_{a=a_h} = \frac{1}{a_h S_h}
\eeq
so
\barr
\overset{\bm .}{a} = \frac{d(a/a_h)}{d(t/t_h)} \frac{a_h}{t_h} =  \frac{d(a/a_h)}{d(t/t_h)} a_h^2 S_h .
\earr
Eq.~(\ref{Equation:OmegaGW0}) can then be written
\barr
\Omega_{\rm GW}\rho_c c^2 &=& \frac{32}{5} \frac{G^{7/3}}{c^5} \mathcal{M}^{10/3} (GM_{12})^{1/3} (a_h^2 S_h)^{-1} f \times\nonumber\\
&\times&\int_{t_r(\zmax)}^0 dt_r \, \displaystyle\sum_{a_c<a_n<a_h}\frac{4\pi}{3n}\left(\frac{2\pi f_r}{n}\right)^{5/3} g(n,e(a_n))\,\dot{\cal N}_m (t_h(a_n,t_r)) \left| \frac{d(a/a_h)}{d(t/t_h)}(a_n)\right |^{-1}.
\earr
It is convenient to return to $z$ as the integration variable, in terms of which
\bsub
\barr
\Omega_{\rm GW}\rho_c c^2 &=& \frac{32}{5} \frac{G^{7/3}}{c^5} \mathcal{M}^{10/3} (GM_{12})^{1/3} (a_h^2 S_h)^{-1} f \nonumber\\
&\times&\int_0^{\zmax} \frac{dz}{H(z)(1+z)}\, \displaystyle\sum_{a_c<a_n<a_h}\frac{4\pi}{3n}\left(\frac{2\pi f_r}{n}\right)^{5/3} g(n,e(a_n))\,\dot{\cal N}_m (z_h(a_n,z)) \left| \frac{d(a/a_h)}{d(t/t_h)}(a_n)\right |^{-1} \\
&=& \frac{64\,(2\pi f)^{8/3}}{15\,Gc^5H_0} (GM_{12})^{11/3}\mathcal{Q}^2 (a_h^2 S_h)^{-1} \nonumber\\
&\times&\int_0^{\zmax} \frac{(1+z)^{2/3}}{ \left[\Omega_{\rm M}(1+z)^3+\Omega_{\Lambda}\right]^{1/2}} dz \, \displaystyle\sum_{a_c<a_n<a_h} n^{-8/3} g(n,e(a_n))\,\dot{\cal N}_m (z_h(a_n,z)) \left| \frac{d(a/a_h)}{d(t/t_h)}(a_n)\right |^{-1}
\earr
\esub
where $z_h(a,z)$ is the formation redshift of a binary which has semimajor axis $a$ at redshift $z$. 

According to Eqs. (\ref{Equation:a_h}) and (\ref{Equation:S_h(r_infl)}),
\barr
a_h^2S_h =
\left(\frac{G\mathcal{Q} M_{12}}{4\sigma^2}\right)^2 \times 4\sqrt\frac{GM_{12}}{(GM_{12}/\sigma^2)^5} = 
\frac{1}{4}\mathcal{Q}^2\sigma
\earr
so
\bsub\label{Equation:OmegaGW1}
\barr
\Omega_{\rm GW}\rho_c c^2 &=& \frac{256\,(2\pi f)^{8/3}}{15\,Gc^5H_0} (GM_{12})^{11/3} \sigma^{-1}(M_{12}) \nonumber\\
&\times&\int_0^{\zmax} \frac{(1+z)^{2/3}}{ \left[\Omega_{\rm M}(1+z)^3+\Omega_{\Lambda}\right]^{1/2}} dz \, \displaystyle\sum_{a_c<a_n<a_h} n^{-8/3} g(n,e(a_n))\,\dot{\cal N}_m (z_h(a_n,z)) \left| \frac{d(a/a_h)}{d(t/t_h)}(a_n)\right |^{-1},\\
\sigma(M_{12}) &=&  180 \left(\frac{M_{12}}{10^8 \msun}\right)^{1/5}\,\kms .
\earr
\esub
(The second equation is the $M-\sigma$ relation, as given below in Eq.~\ref{Equation:M-sigma}). We now relax our assumption of a single set of values $\{M_{12}, q, e_0\}$ and generalize Eq.~(\ref{Equation:OmegaGW1}) 
to consider a population of binaries with different initial parameters.
Furthermore we add two parameters related to nuclear rotation: $\theta_0$, the binary's initial inclination, and $\eta$, a parameter that
determines the degree of ordered rotation of a nucleus; both parameters are defined and discussed in more detail in
\S\ref{Section:nuc-rotation}.
We redefine the merger rate as the rate per unit comoving volume, and per unit of $M_{12}$, $q$, $e_0$, $\theta_0$ and $\eta$.
The resulting expression is
\barr
\Omega_{\rm GW}\rho_c c^2 &=& \frac{256\,(2\pi f)^{8/3}}{15\,Gc^5H_0} 
\int_{1/2}^1 d\eta \int_{\qmin}^1 dq \int_{\mmin}^{\mmax} dM_{12} \int_{0}^{\pi} d\theta_0 \int_{0}^{1} de_0 \int_0^{\zmax} dz \,(GM_{12})^{11/3} \sigma^{-1}(M_{12}) \nonumber\\
&\times& \frac{(1+z)^{2/3}}{ \left[\Omega_{\rm M}(1+z)^3+\Omega_{\Lambda}\right]^{1/2}} \nonumber\\
&\times&\displaystyle\sum_{a_c<a_n<a_h} n^{-8/3} g(n,e(a_n))\,\dot{\cal N}_m (z_h(a_n,z),M_{12},q,e_0,\theta_0,\eta) \left| \frac{d(a/a_h)}{d(t/t_h)}(a_n)\right |^{-1} .
\earr
Note our additional assumption that host galaxy properties, such as $\sigma$, are determined by $M_{12}$. 
The characteristic strain, $h_c(f)$, is given in terms of $\Omega_\mathrm{GW}(f)$ by Eq.~(\ref{Equation:OmegaGW}),
so that
\barr\label{Equation:hcfinal}
h_c(f) &=& \Bigg[  
\frac{1024\pi\,(2\pi f)^{2/3}}{15c^7H_0} 
\int_{1/2}^1 d\eta \int_{\qmin}^1 dq \int_{\mmin}^{\mmax} dM_{12} \int_{0}^{\pi} d\theta_0 \int_{0}^{1} de_0 \int_0^{\zmax} dz \,(GM_{12})^{11/3} \sigma^{-1}(M_{12})\, \nonumber\\
&\times& \frac{(1+z)^{2/3}}{ \left[\Omega_{\rm M}(1+z)^3+\Omega_{\Lambda}\right]^{1/2}} \nonumber\\
&\times&\displaystyle\sum_{a_c<a_n<a_h} n^{-8/3} g(n,e(a_n))\,\dot{\cal N}_m (z_h(a_n,z),M_{12},q,e_0,\theta_0,\eta) \left| \frac{d(a/a_h)}{d(t/t_h)}(a_n)\right |^{-1}\Bigg]^{1/2}.
\earr

\begin{figure*}[h!]
	\centering
	\subfigure{\includegraphics[width=0.49\textwidth]{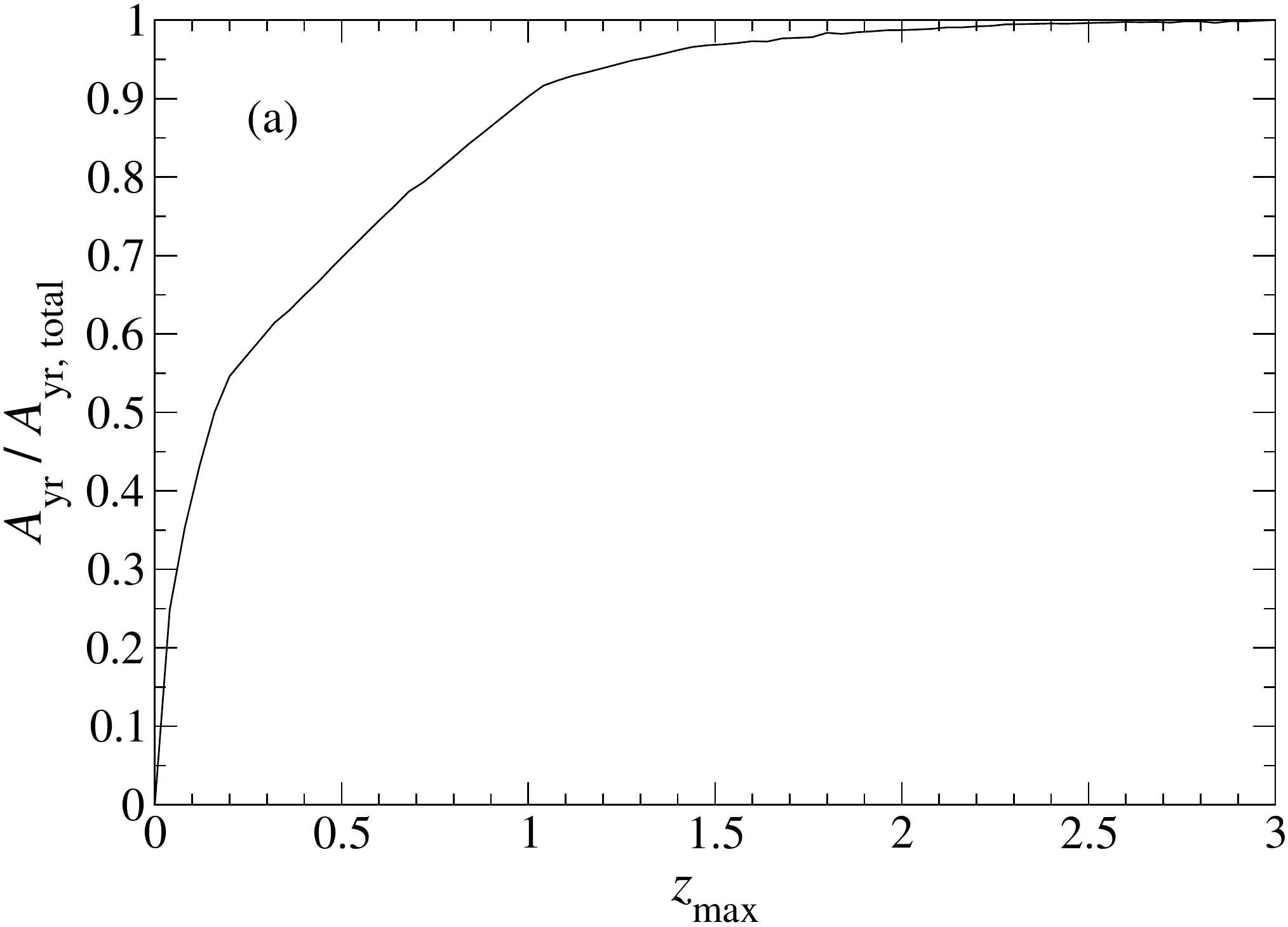}}
	\subfigure{\includegraphics[width=0.49\textwidth]{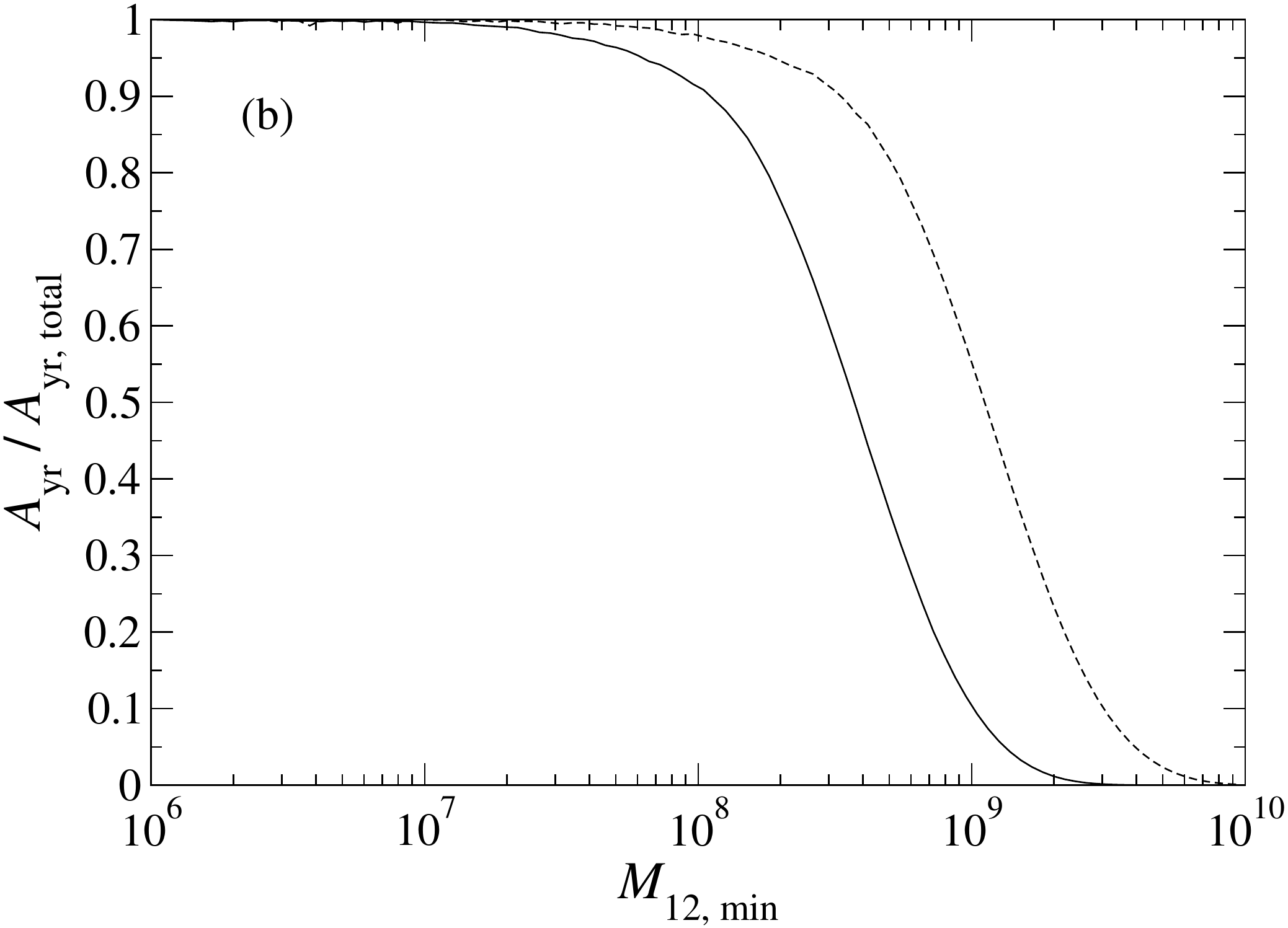}}
	\subfigure{\includegraphics[width=0.49\textwidth]{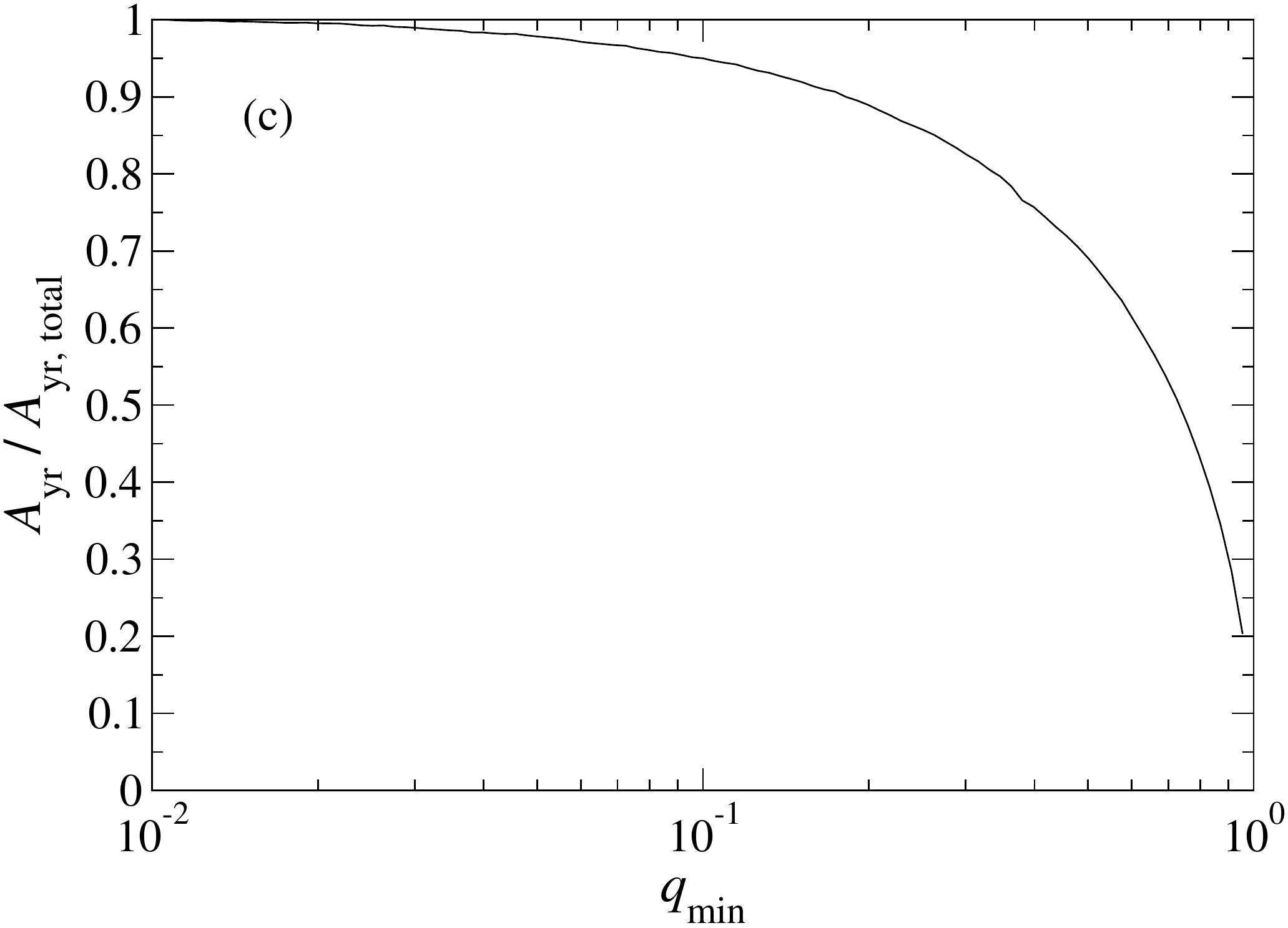}}
	\subfigure{\includegraphics[width=0.49\textwidth]{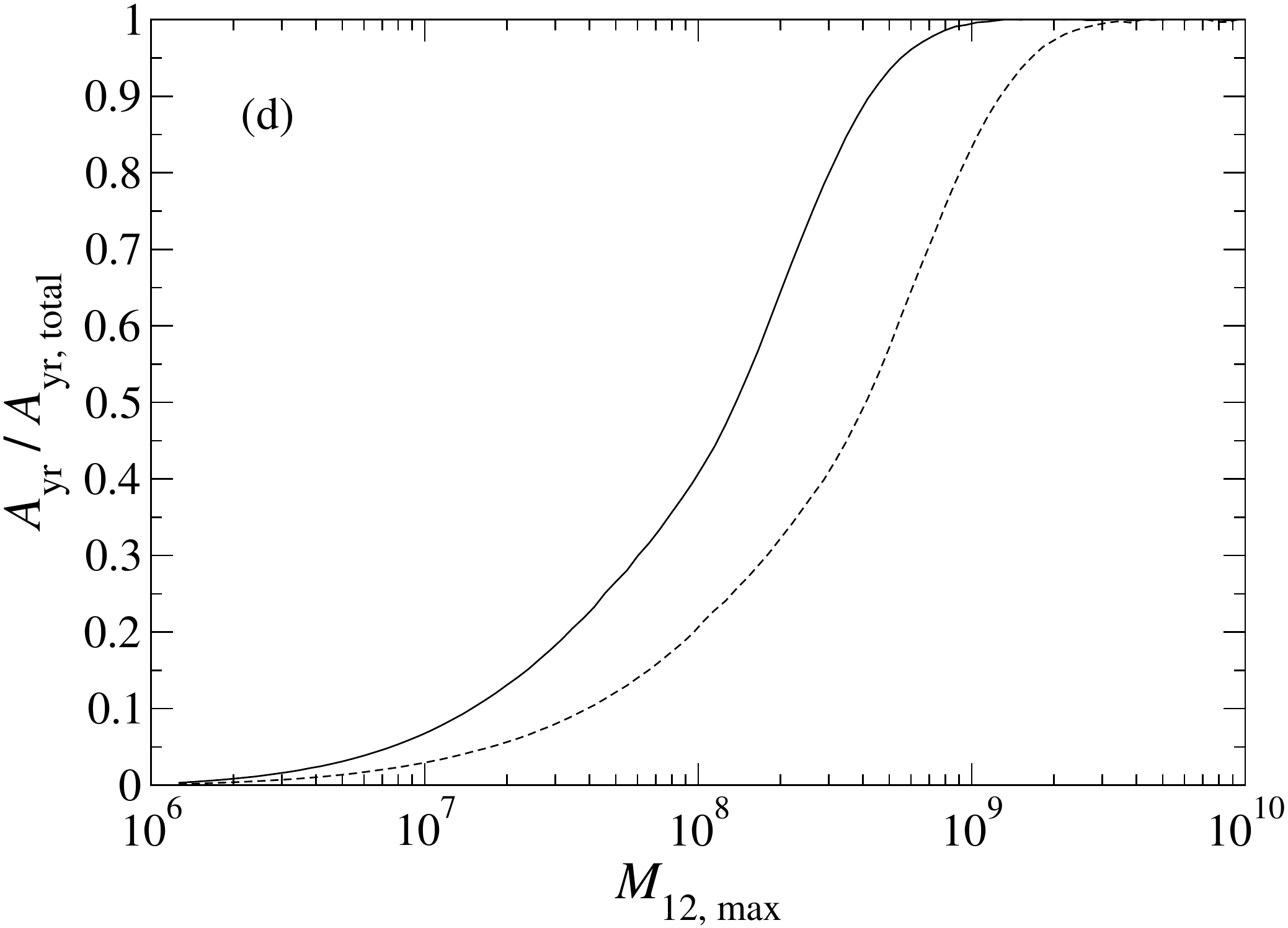}}
\caption{
Fraction of the total GW strain  at $f=1\mathrm{yr^{-1}}$ contributed by massive binaries with 
(a) $z<\zmax$, (b)  $M_{12}>\mmin$, (c) $q>\qmin$ or (d) $M_{12}<\mmax$. If not otherwise specified, $\zmax=4$, $\qmin=1/100$, $\mmin=10^6\msun$ and $\mmax=10^{10}\msun$. Plots assume circular-orbit binaries and \sbh-bulge mass ratio $\beta=0.001$ (straight lines) or $\beta=0.003$ (dashed lines); for (a) and (c) both lines look identical.
}
\label{Figure:zmax}
\end{figure*}

In what follows, we adopt the following limits on the integrals that appear in Eq.~(\ref{Equation:hcfinal}): 
$\zmax=3$, $\qmin=1/10$, $\mmin=10^6\msun$, $\mmax=10^{10}\msun$. 
As Fig.~\ref{Figure:zmax} shows, in this way, we account for more than 95\% of the total signal.

\subsection{Galaxy merger rate}
\label{Section:mergerRate}

Galaxy merger rates are customarily expressed ``per galaxy'', i.e., $\dot N_{\rm mergers}(M_{\rm gal},z)$ is the rate at which
a ``primary'' galaxy, of mass $M_{\rm gal}$, experiences mergers with other galaxies at redshift $z$.
We convert such an expression into the merger rate per unit galaxy mass, per unit mass ratio, by multiplying it by 
(i) the galaxy mass function at a given redshift $\phi(M_{\rm gal},z)$; 
and (ii) the distribution of galaxy mass ratios $q_{\rm gal}$, which we assume to be $\propto 1/q_{\rm gal}$ following \cite{Lopez2012} and \cite{Sesana2013a}:
\barr
\dot{\cal N}_m \, dq\, dM_{12}\, de_0\, d\theta_0 &=& \frac{dN_{\rm mergers}}{dt}(\mgal,z) \frac{1}{q_{\rm gal}(-\ln{q_{\rm lim}})}
\phi(\mgal,z) \frac{d\mgal}{dM_{12}} \mathcal{F}(e_0,\theta_0,\eta) \, dq \,dM_{12} \,de_0 \,d\theta_0 .
\label{Equation:NdotMergers}
\earr
Here $q_{\rm lim}$ is the minimum mass ratio selected in counting galaxy pairs and 
$\mathcal{F}(e_0,\theta_0,\eta)$ is the joint distribution of binary initial parameters $e_0$, $\theta_0$ and of $\eta$. 
Unfortunately, little is known about $\mathcal{F}$.
With regard to its dependence on $M_{12}$, $q$ and $z$, we take a ``maximally-uninformed'' stance and 
posit no dependence.

In what follows, we consider the following possibilities for the dependence of 
$\mathcal{F}$ on $e_0$, $\theta_0$ and $\eta$. 
In each case we normalize $\mathcal{F}$ so that
\barr
\int_{1/2}^1\int_0^1\int_0^\pi \mathcal{F}(e_0,\theta_0,\eta)\, d\theta_0 \,de_0 \,d\eta = 1 .  \nonumber
\earr

\begin{enumerate}
\item 
$e_0$, $\theta_0$ and $\eta$ are the same for all binaries:
\barr
\mathcal{F}\, d\theta_0 \,de_0 \,d\eta= \delta(e_0-e_{0,1})\,\delta(\theta_0-\theta_{0,1})\,\delta(\eta-\eta_1)\,de_0\,d\theta_0\,d\eta .
\earr
This includes the simplest case in which all binaries are initially circular, $e_{0,1}=0$; 
in this special case, the distributions over $\theta_0$ and $\eta$ do not matter because a circular binary remains circular in the course of its evolution.
\item $e_0$ has a ``thermal'' distribution, $dN/de_0=2e_0$, 
and the binary is either corotating from the beginning ($\theta_0=0$) or its orbital plane has no preferred initial direction:
\bsub\label{Equation:e_0-thermal}
\barr
\mathcal{F}\, d\theta_0 \,de_0 &=& \mathcal{F}_1(\eta)\cdot 2e_0\,de_0\,\delta(\theta_0)\,d\theta_0,\,\mathrm{or}\\
\mathcal{F}\, d\theta_0 \,de_0 &=& \mathcal{F}_1(\eta)\cdot e_0\,de_0\,\sin{\theta_0}\,d\theta_0.
\earr
\esub
The functional form of $\mathcal{F}_1(\eta)$ is discussed in Section~\ref{Section:nuc-rotation}. 
\end{enumerate}

For the galaxy (stellar) mass function $\phi(M_{\rm gal},z)$, we adopt the analytic expressions of 
\cite{Thanjavur2016} (Schechter fit) for $z\leq0.2$ and \cite{Ilbert2013} (double Schechter fit) for $z>0.2$. With the possible exception of the highest galaxy masses, both expressions agree well with determinations by other groups (see Fig. 7 in \cite{Thanjavur2016} and Fig. 6 in \cite{Tomczak2014}). 
The galaxy merger rate and the relation between galaxy mass and \sbh\ mass we use are discussed in the next two subsections.

As for the merger rate per galaxy, we adopt the analytic expression of \cite{Xu2012}:
\barr
\frac{dN_{\rm mergers}}{dt} = 0.053\,\mathrm{Gyr^{-1}}\left(\frac{M_{\rm gal}}{10^{10.7}\msun}\right)^{0.3}\frac{(1+z)^{2.2}}{1+z/8}.
\label{Equation:Xu}
\earr

\noindent
\citet{Xu2012} obtained this expression by dividing their observed galaxy pair fraction by the average major merger timescale ($\sim 300\,\mathrm{Myr}$) taken from \citet{merger2}, who performed hydrodynamical simulations of disk galaxy mergers for a number of different masses, mass ratios and initial orbits.\footnote{The simulations assume that observable galaxies are surrounding by dark-matter halos; otherwise merging timescales would be much longer.} 

Given that the galaxy merger process is very complicated, there are undoubtedly many uncertain factors that influence the merger rate, both through the (observed) pair fraction and the (simulated) merger timescales. As shown in Figure 15 of \cite{Robotham2014}, the pair-fraction estimates of \citet{Xu2012} are consistent with the average value of the other studies, although a factor of $\sim 2$ discrepancy between different papers exists. Some of these discrepancies might be caused by  differences in photometric completeness levels and blending issues in different wavebands used to identify mergers \cite{differentBands}.

As noted above, we equate the galaxy merger rate, Eqs. (\ref{Equation:NdotMergers}) and (\ref{Equation:Xu}), with the rate at which ``hard'' \sbh\ binaries are forming -- in other words, we have ignored the time for two \sbhs\ to reach the center of the merger remnant. As  shown by \citet{Merritt2000} and \citet{Dosopoulou2016}, this time can indeed be long for very low mass ratios $q$ and $q_\mathrm{gal}$, approaching the Hubble time for $q\lesssim10^{-3}$.
However, for binaries of any mass with $q>0.1$ (which contribute almost all of the GW background, see Fig.~\ref{Figure:zmax}c) this time is always shorter than $100\,\mathrm{Myr}$ and can be ignored. 

The merger timescales from \cite{merger2} are actually ``observability timescales'' -- the total amount of time a merging pair of galaxies spends at a certain observable merger stage; for example, in the case of \cite{Xu2012} the observability criterion is a projected separation between 5 and 20 $h^{-1}\,\mathrm{kpc}$, and the timescales from \cite{merger2} are chosen accordingly. This way, the uncertainty in the definitions of the beginning and the end of a merger is eliminated. There are, however, a number of other caveats in this approach:

\begin{enumerate}
\item In all of their simulations, Lotz et al. \cite{merger2} assume both galaxies to be disklike while the galaxies contributing most of the GW signal are elliptical (Fig.~\ref{Figure:h_c(galaxy_type)}a). 
The dependence of merger timescale on galaxy morphology might be one of the largest sources of uncertainty.
\item We use the same average timescale for all  mergers of the same mass and redshift, ignoring the dependence on various parameters of a galaxy pair that could be correlated with GW emission, such as mass ratio or initial spin orientations of the galaxies. Lotz et al. found that orientations have little effect for close pairs observed at $5 < r_p < 20 \,h^{-1}\,\mathrm{kpc}$. On the other hand, timescales for equal-mass mergers could be $\sim30\%$ longer compared to $q=1/3$. Since binaries with higher $q$ contribute more signal, that could lead us to underestimate the merger timescale and, consequently, overestimate $h_c$, but only by $\lesssim15\%$ since $h_c\propto\sqrt{\dot{N}_{\rm mergers}}$.
\item Merger timescales depend on gas fraction: for equal-mass mergers, they become considerably shorter when gas fractions are high \cite{merger2-gas}. 
The reason is that the disk galaxies having higher gas fractions are harder to deblend at close separations, which reduces the time interval over which two galaxies can be observed as a close pair.
However, this effect vanishes for $q\lesssim1/3$. \citet{Lotz2011} have calculated the average timescale for $1/4<q<1$ and three different assumptions about gas fraction and found it to be almost the same as the value we use ($330\,\mathrm{Myr}$) for all three cases with a weak dependence on redshift (at least at $z\lesssim1$). 
\item The hydrodynamical simulations of Lotz et al. give a significantly lower merger timescale estimate than the semi-analytical model of \citet{merger1}, who used a mock galaxy
catalog derived from the Millennium simulation. Kitzbichler \& White assume that the secondary galaxy moves in a circular orbit inside the constant potential of the primary, and this is probably not a good approximation for close pairs of nearly equal mass.
\item All of the simulation timescales adopted here assume highly eccentric orbits of galaxies with pericenter distances $\sim0.01-0.05$ times the virial radii of the progenitors. However, galaxies merging on circular orbits or with larger impact parameters can be identified as close pairs for $15-40\%$ longer \cite{merger2-r_p}. 
\item Also, at higher redshifts the mergers could proceed faster for the reason that the galaxy sizes for a given mass are smaller. According to \citet{galaxySize}, the characteristic radii of big ellipticals are $\sim2$ times smaller at $z=1$ compared to $z=0$.
\item Finally, all of the galaxy merger simulations posit that the observed, luminous parts of galaxies are surrounded by extensive, dynamically-active, dark-matter haloes. If the dark-matter haloes are not present, or if the ``dark matter'' is not particle in nature, merger times would be much longer; indeed most of the observed interacting pairs would ``pass in the night'' and never merge
\cite{Toomre1977}. It is not our intention here to stake out a position in the dark-matter debate \citep{McGaugh2015}. But we do note the troubling lack of corroborative evidence for the merger hypothesis, and the fact that some observationally-based studies reach conclusions contradicting the predictions of $\Lambda$CDM cosmology, such as overabundance of bulgeless giant galaxies \cite{Kormendy2010,Weinzirl2009} or that giant ellipticals couldn't have been formed from disk galaxy mergers \cite{NaabOstriker2009}.
\end{enumerate}
\noindent
The effects on $h_c(f)$ of systematic uncertainties in the merger rate are discussed in more detail in Section~\ref{Section:Results} (Eq.~\ref{Equation:A,f_b}).


\subsection{\sbh\ demographics}
\label{Section:mbh-mbulge}

We assume a strict proportionality between SBH mass and the mass of the stellar bulge, defining the parameter
\barr
\beta \equiv M_{\rm BH} / M_{\rm bulge} .
\earr
Estimates of this quantity have evolved over time; we identify three epochs.

\begin{enumerate}
\item Kinematical modeling of early-type galaxies by Magorrian et al. (1998) \citep{Magorrian1998} favored a high value, $\beta \approx 0.006$. This value was immediately seen to be inconsistent with (i.e. larger than) the mean mass ratio in active galaxies, either as predicted by the Soltan argument, or as estimated via reverberation mapping in individual galaxies \citep{Richstone1998}. \item In 2000, the $M-\sigma$ relation was discovered by restricting the sample to galaxies with clear, prima facie evidence for a Keplerian velocity rise, leading to a much lower estimate, $\beta \approx 0.0012$. This smaller value eliminated the discrepancies with the other two methods \citep{MerrittFerrarese2001}. 
\item Starting around 2006, and continuing until the present day, most authors have sought to be comprehensive, including in their samples essentially every published \sbh\ mass without regard to the presence or absence of a kinematical signal. These studies (as summarized by \citep{Kormendy2013}) find a larger value, $\beta \approx 0.003$, that is once again inconsistent with (i.e. larger than) \sbh\ masses in AGN \citep{Reines2015}.
This is the value of $\beta$ assumed in all recent calculations of $h_c(f)$ \citep{Sesana2013a,McWilliams2014}.
\end{enumerate}

\noindent
We note here a worrisome phenomenon: when stellar (as opposed to gas) kinematical data for a galaxy are re-modeled independently, the results for $M_{\rm BH}$ are often very different than in the ``discovery'' paper: the best-fit \sbh\ mass is found to be much lower; there is a range of equally-likely masses; or only an upper limit can be established. 
A recent example is NGC 1277, where claims of a $\sim2\times10^{10}\msun$ \sbh\ 
\citep{vandenBosch2012} were subsequently found to be too large by factors of 3-5 \citep{Emsellem2013,Walsh2016}.
Indeed, beyond the Local Group, few if any galaxies show evidence for a central increase in the rms {\it stellar} velocities on the relevant spatial scales (see Figure 2.5 in \citep{DEGN}). 
Brightest cluster galaxies (BCGs), which are strongly represented among galaxies with ``overmassive'' \sbhs\ \citep{Volonteri2013}, are particularly difficult cases due to their low central densities, so that stellar velocity dispersions measured near the projected center are strongly weighted by stars that are far from the \sbh. 
For instance, M87, the BCG in the Virgo Cluster, exhibits no prima facie evidence for a central \sbh\ in the stellar velocities, and the value of $M_{\rm BH}$ derived from the stellar data in M87 depends critically on what mass-to-light ratio is assigned to the stars
\citep{Gebhardt2009,Gebhardt2011}; 
furthermore the value of $M_{\rm BH}$ derived from the stellar data is a factor $\sim 2$ greater than
the value derived from the gaseous rotation curve \citep{Walsh2013}.
(The latter {\it does} exhibit a clear Keplerian rise and the value of $M_{\rm BH}$ derived from it is much more robust, having remained
essentially constant since its first determination in 1997 \citep{Macchetto1997}.)
In effect, what is being measured in such galaxies is not the \sbh\ mass, but rather the mass of the \sbh\ plus the mass of the stars within some region the size of which is comparable to the resolution limit set by the telescope and which may be much larger than the \sbh\ influence radius. Disentangling the two contributions can be extremely difficult \citep{VME2004}. Claims that the \sbh\ influence radius has been ``resolved'' in such galaxies are always suspect, since they are based -- not on an observed Keplerian velocity rise -- but rather on the {\it assumption} that the influence radius is given by $\sim GM_{\rm BH, est}/\sigma^2$ with $M_{\rm BH,est}$ the {\it estimated} \sbh\ mass.
It is axiomatic that a ``best-fit'' value of $M_\mathrm{BH}$ will have an influence radius larger than the instrumental resolution,
whether or not the data from which $M_{\rm BH, est}$ was derived contain any useful information about the presence of a central
mass concentration.

\begin{figure}[h!]
\includegraphics[width=0.79\textwidth]{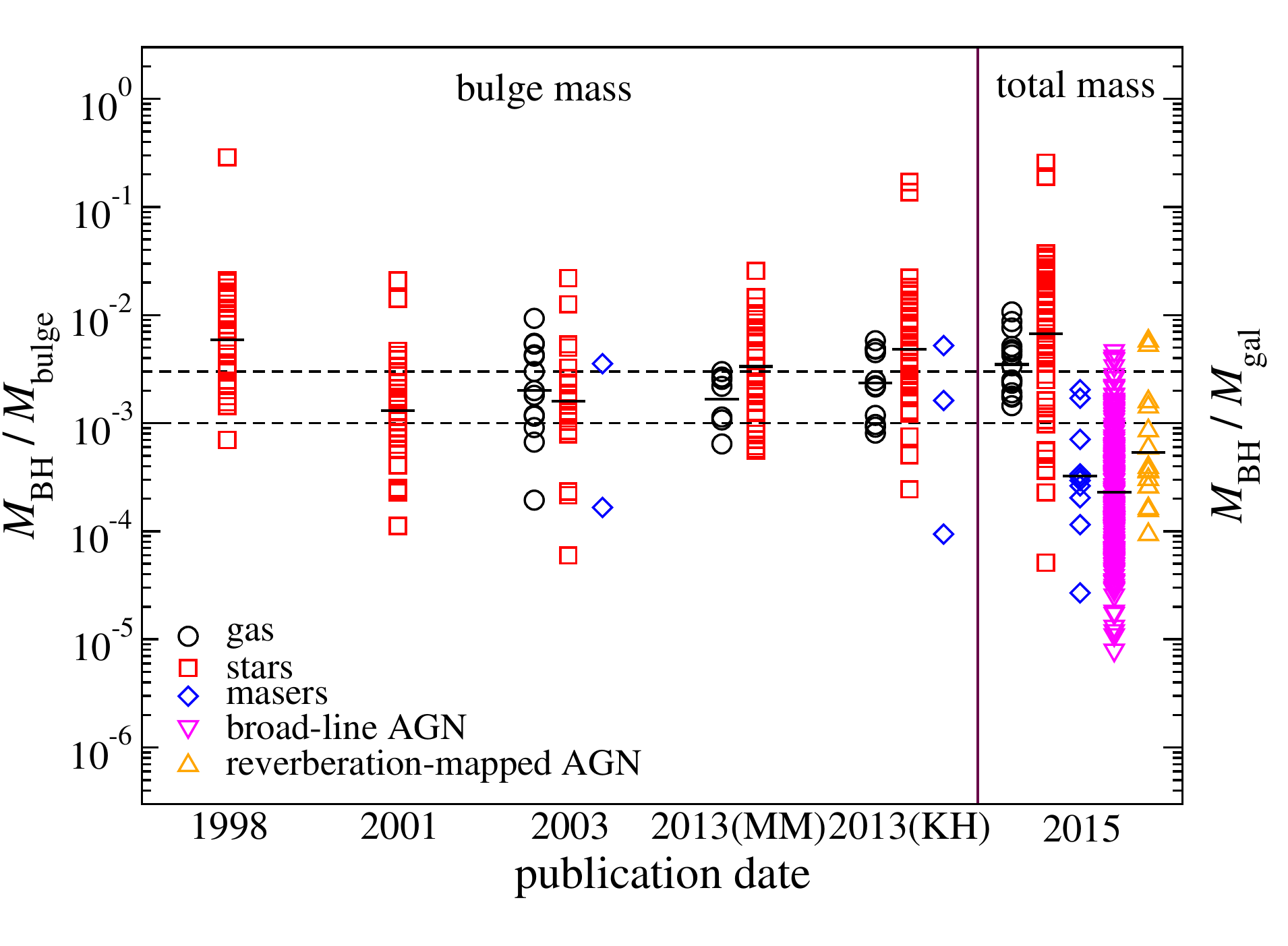}
\caption{Published estimates of the ratio $M_\mathrm{BH}/M_\mathrm{galaxy}$ 
(or $M_\mathrm{BH}/M_\mathrm{bulge}$)
ordered by publication date and \sbh\ mass measurement method used.
Every point corresponds to a single galaxy; mean values are indicated with horizontal ticks. 
Horizontal dashed lines mark the values 0.003 (the currently accepted value) and 0.001 (the more 
conservative estimate considered in this paper).
References: Magorrian et al. (1998) \citep{Magorrian1998}, Merritt \& Ferrarese (2001) \citep{MerrittFerrarese2001},
Marconi \& Hunt (2003) \citep{Marconi2003}, McConnell \& Ma (2013) \citep{McConnellMa2013}, Kormendy \& Ho(2013) 
\citep{Kormendy2013}, Reines \& Volonteri (2015) \citep{Reines2015}.
}
\label{Figure:M_BH-M_bulge}
\end{figure}

Figure~\ref{Figure:M_BH-M_bulge} presents a compilation from the literature of estimates of 
$M_\mathrm{BH}/M_\mathrm{bulge}$.
In line with the discussion in the previous paragraph, we make the following observations.
i) Estimates of the mean $M_\mathrm{BH}/M_\mathrm{bulge}$ reached a minimum near 2000,
following the winnowing of the stellar-based $M_\mathrm{BH}$ values by \citet{FerrareseMerritt2000}.
Around this time, estimates based on stellar and gas data were consistent.
ii) Estimates of $M_\mathrm{BH}/M_\mathrm{bulge}$ 
made since that time have crept back upward, particularly in the case of the stellar-based masses,
and particularly in the most-massive galaxies (BCGs).
iii) In any given study, the ordering of $\langle M_\mathrm{BH}/M_\mathrm{bulge}\rangle$ 
typically obeys stars $>$ gas $>$ AGN.
This is consistent with the fact that the stellar data rarely exhibit a Keplerian rise, hence the $M_\mathrm{BH}$ 
values are likely to be biased upward, as discussed above.
Estimates of $M_\mathrm{BH}$ in AGN, at the other extreme, use measured velocities of broad-emission-line gas that lies 
well inside the \sbh\ influence sphere, 
hence is guaranteed to be responding almost entirely to the gravitational force of the \sbh.

So far we have emphasized uncertainties in $M_\mathrm{BH}$.
Table \ref{Table:BulgeMasses} points out a completely independent source of worry.
Estimates of $M_\mathrm{bulge}$ in a given galaxy can exhibit wide variation from author to author.
We are unable to give a reason for this, except to note that different authors base their estimates
on luminosities measured in different passbands, carry out the bulge-disk decompositions differently,
and make different assumptions about the stellar IMF and/or the mass-to-light ratio.
(\citet{Marconi2003} estimate bulge masses using the virial theorem, not measured luminosities.)

\begin{table}[h!]
\caption{Bulge mass estimates (Solar masses)}
\centering
\begin{tabular}{l c c c}
\hline\hline
Reference & M87 & NGC4459 & NGC3377 \\ [0.5ex] 
\hline
\citet[][2003]{Marconi2003} & $6.2\times 10^{11}$ & $3.6\times 10^{11}$ & $7.8\times 10^{10}$  \\
\citet[][2013]{McConnellMa2013} & $1.3\times 10^{12}$ & --- & $2.4 \times 10^{10}$  \\
\citet[][2013]{Scott2013}& $2.3\times 10^{11}$ & $2.0\times 10^{10}$ & $2.0 \times 10^{10}$ \\
\citet[][2013]{Kormendy2013} & $5.3\times 10^{11}$ & $7.6\times 10^{10}$ & $3.2\times 10^{10}$ \\
\citet[][2015]{Reines2015} & $2.4\times 10^{11}$ & $3.6 \times 10^{10}$ & $1.4 \times 10^{10}$ \\ 
\citet[][2016]{Savorgnan2016} & $2.6\times 10^{11}$ & $2.9 \times 10^{10}$ & $4.0 \times 10^{10}$ \\ [1ex]
\hline
\end{tabular}
\label{Table:BulgeMasses}
\end{table}

Based on these arguments, we adopt $\beta\approx  0.001$ (the value in 2001) as our preferred estimate of this ratio.
We note that such a value is lower than what previous authors have assumed when estimating the stochastic GW background and thus implies a lower PTA signal than in the earlier studies 
(all else being equal).
But given the sources of uncertainty discussed above, we  present results for other (higher) values of $\beta$ as well in what follows.


As for the fraction of mass of the galaxy contained in the bulge $f_{\rm bulge}$, we adopt the prescription of \citet{Simon2016}: for quiescent (elliptical) galaxies $f_{\rm bulge}=0.9$ for $\mgal > 10^{11}\msun$, declining log-linearly to $f_{\rm bulge}=0.25$ at $\mgal = 10^{10}\msun$, and for all star-forming (spiral) galaxies $f_{\rm bulge}=0.25$. 
We do not allow for any scatter in these parameters which makes our predictions for $h_c(f)$ somewhat lower than those of \cite{Sesana2013a} or \cite{Simon2016}. 

Given such large discrepancies between different aluthors and different methods, we consider most of  the observed scatter in the $M-\sigma$ relation to be caused by measurement errors and ignore any possible intrinsic scatter. That makes our predictions for $h_c(f)$ somewhat lower than those of \cite{Sesana2013a} or \cite{Simon2016}. Figure 7 of \cite{Simon2016} shows exactly how much we would underestimate the GW amplitude given intrinsic scatter; for example, a value of 0.3--0.4 dex reported by \cite{McConnellMa2013} and \cite{Kormendy2013} implies a factor of $\sim1.5$ difference in amplitude compared to zero scatter.  

\subsection{Nuclear rotation}
\label{Section:nuc-rotation}

In the simple galaxy models adopted here, rotation is implemented by supposing that some fraction of the stars on any
given orbit have had the direction of their orbital angular momentum flipped compared with a nonrotating (isotropic) model.
The parameter $\eta$ is defined as the fraction of stars having a positive angular momentum component along the assumed
axis of rotation; $\eta=1/2$ corresponds to a nonrotating galaxy, $\eta=1$ to a maximally-rotating one. 

\citet{Sesana2014} present a compilation from the literature of values of $V/\sigma$: the ratio of mean (streaming) velocity
to velocity dispersion.
They find that the following functions (normalized here to unit total number)
are good representations of the observed distribution of $x\equiv V/\sigma$ 
for elliptical galaxies and for the bulges of spiral galaxies respectively:
\bsub\label{Equation:vsigmasesana}
\begin{eqnarray}
&&\mathrm{Ellipticals:} \ \ \ \ N(x)dx = 4.27\; \exp{\left(-3.98\; x^{0.96}\right)} dx \\
&&\mathrm{Bulges:} \ \ \ \ N(x)dx = 2.18\; x^{0.24} \exp{\{-0.5\left[\left(x-0.47\right)/0.24\right]^2\}} dx .
\end{eqnarray}
\esub
Both functions are effectively zero for $V/\sigma>1$.

\begin{figure}[h!]
\includegraphics[width=0.49\textwidth]{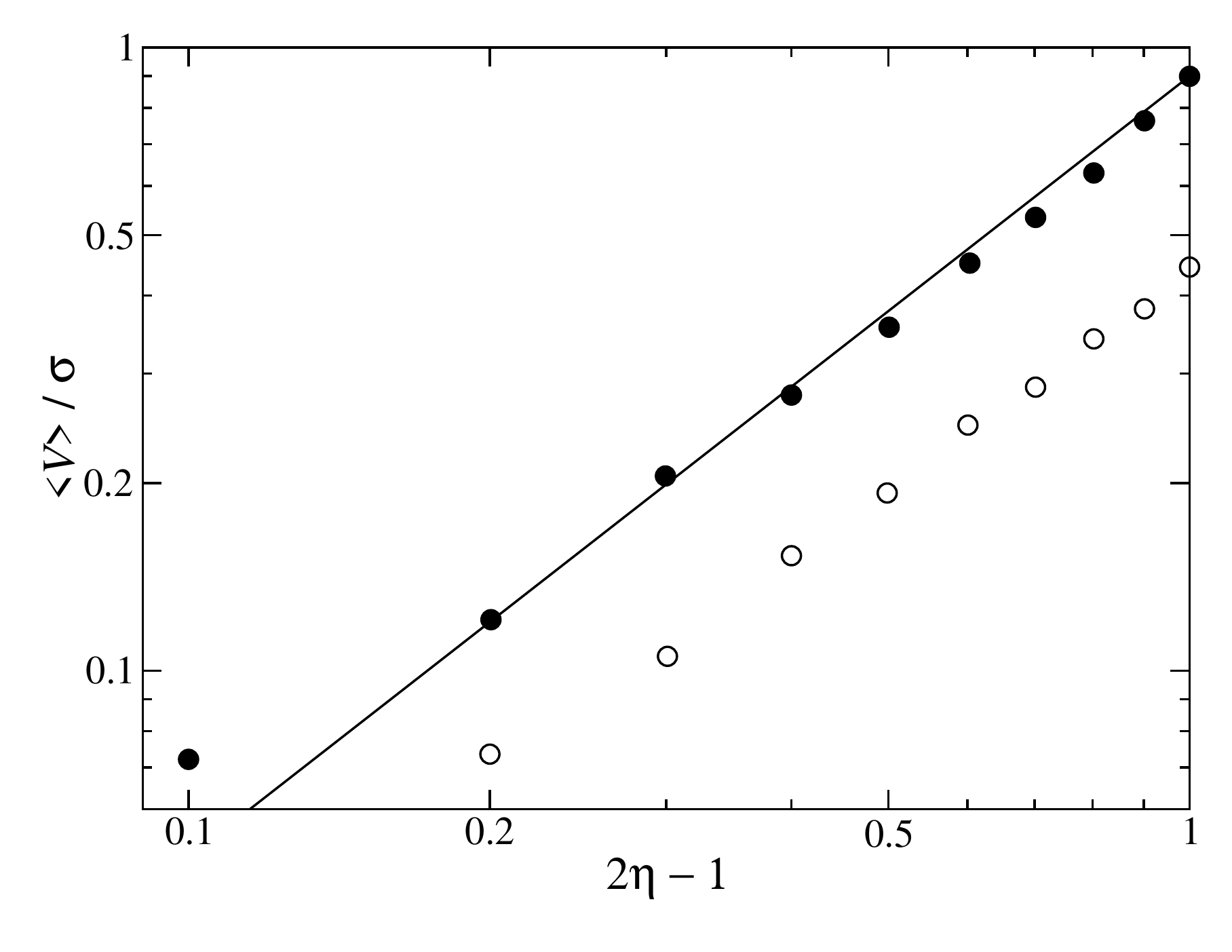}
\includegraphics[width=0.49\textwidth]{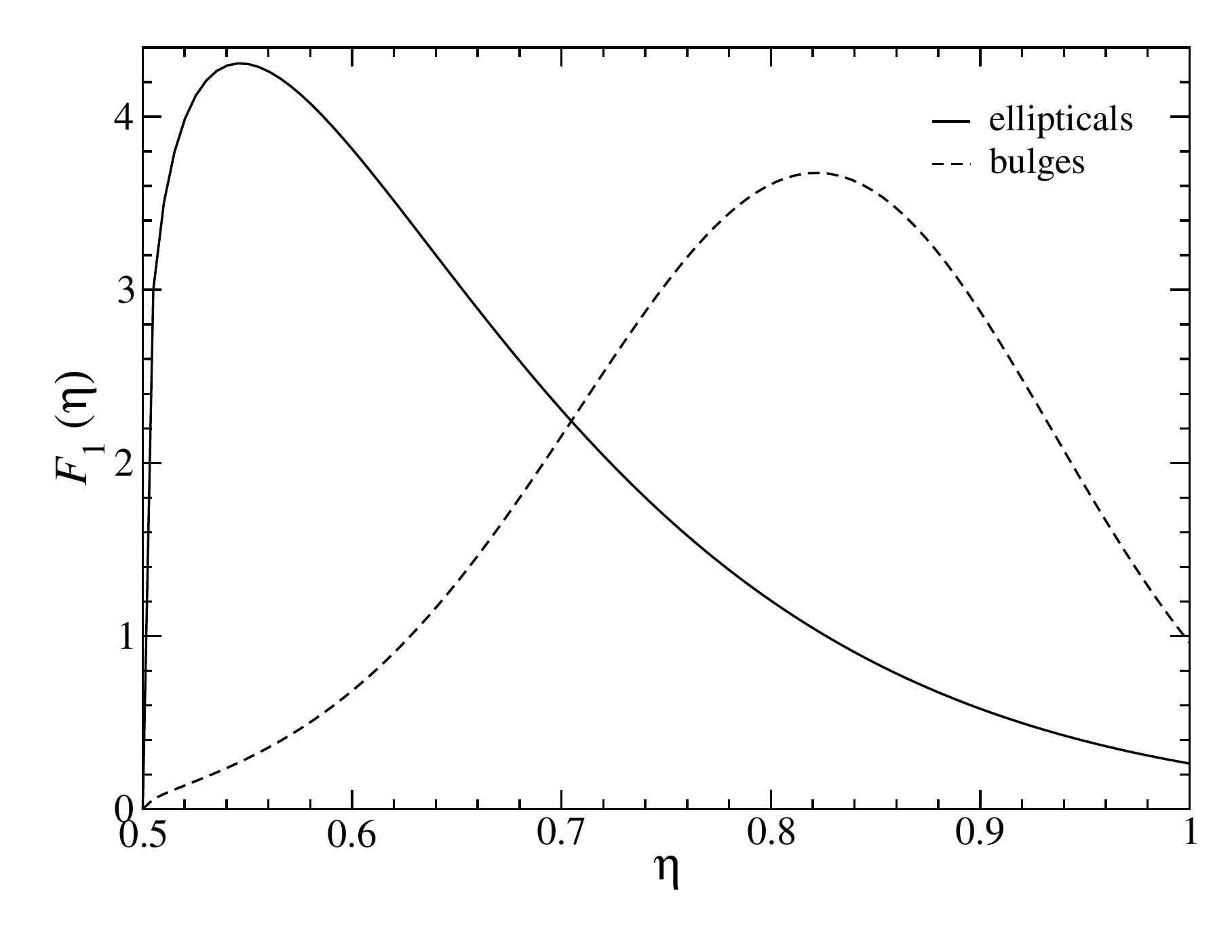}
\caption{Left: rotational properties of the galaxy models as a function of the parameter $\eta$. Filled and open circles show
values measured at the galaxy half-mass radius and the \sbh\ influence radius respectively, the line is Eq.~(\ref{Equation:vsigmaeta}).
The nonrotating model from which the rotating models were generated by orbit-flipping was 
described by Dehnen's \cite{Dehnen1993} density law with $\gamma=1$ and with an assumed
\sbh\ mass of $0.002 M_\mathrm{gal}$. Right: two distributions of $\eta$ used in this paper (see Eq.\ref{Equation:F1(eta)} and \ref{Equation:vsigmasesana}).}
\label{Figure:vseta}
\end{figure}

In order to map $\eta$ onto the observed $V/\sigma$, we investigated the observable properties of our models.
Monte-Carlo representations were constructed and the projected, line-of-sight mean velocity and velocity dispersion
were computed along the equatorial plane.
Figure~\ref{Figure:vseta} (left) shows $\langle V\rangle/\sigma$ computed at two projected radii: 
$R=r_\mathrm{infl}$, the \sbh\ influence radius; and $R=R_\mathrm{eff}$, the effective (projected half-mass) radius.
The latter quantity, which is most directly comparable to the $V/\sigma$ values tabulated by \citet{Sesana2014},
is well described by
\beq\label{Equation:vsigmaeta}
\left|\frac{\langle V\rangle}{\sigma}\right|_{R=R_\mathrm{eff}} \approx 0.9 \left(2\eta -1\right)^{1.25} .
\eeq
We used Eq.~(\ref{Equation:vsigmaeta}) to express the relations (\ref{Equation:vsigmasesana}) in terms of $\eta$, so that the $\eta$ part of the merger rate distribution function (Eqs.~\ref{Equation:e_0-thermal}) is
\beq\label{Equation:F1(eta)}
\mathcal{F}_1(\eta) = N(x) \frac{dx}{d\eta}, \ \ \ \ x=0.9 \left(2\eta -1\right)^{1.25} \in \left[0,0.9\right]
\eeq
This function is shown in Fig.~\ref{Figure:vseta} (right).
\subsection{Dynamical evolution of the binary}
\label{Section:Evolution}

Interaction of a massive binary with stars in a galactic nucleus causes changes in the binary's semimajor axis $a$
and eccentricity $e$, as well as its orbital plane; the latter is characterized by the angle $\theta$ between the binary's angular momentum vector and the rotation axis of the nuclear cluster (the latter assumed fixed). 
The only stars we take into account are the ones initially unbound to the binary but with a small enough 
pericenter distance (of the order of the binary semimajor axis $a$) that they can experience a close interaction with it.
The equations describing the coupled changes in $(a,e,\theta)$ can be written \citep{PaperI}
\bsub
\label{Equation:Rate1}
\barr
\frac{da}{dt} &=& \left(\frac{da}{dt}\right)_\star + \left(\frac{da}{dt}\right)_{\rm GR} = 
- a^2 S - \frac{64}{5}\frac{\mathcal{Q} G^3M_{12}^3}{c^5a^3}\; F(e) , \\
\frac{de}{dt} &=& \left(\frac{de}{dt}\right)_\star + \left(\frac{de}{dt}\right)_{\rm GR} = 
a K S - \frac{304}{15}\frac{\mathcal{Q} G^3M_{12}^3}{c^5a^4}\; G(e) , \\
\frac{d\theta}{dt} &=& \left(\frac{d\theta}{dt}\right)_\star = a D S , \\
F(e) &=& \frac{1+(73/24)e^2+(37/96)e^4}{(1-e^2)^{7/2}} , \\
G(e) &=& e\frac{1+(121/304)e^2}{(1-e^2)^{5/2}}
\earr
\esub
where $S\equiv (d/dt)(1/a)$ is the binary hardening rate defined above, 
and $K$ and $D$ are dimensionless rate coefficients.\footnote{Expressed in terms of quantities defined in
 \citep{PaperI}, $D = D_{\theta,1}/H$.}
All three rate equations may depend on $a$, $e$ and $\theta$ as well as on $\eta$ and $q$. 
In deriving the rate coefficients,  averages were taken over the binary argument of periapsis, $\omega$, whose evolution
is ignored. 
The binary's nodal angle $\Omega$ can be expected to evolve in a deterministic way; however we ignore that 
evolution here since it  does not affect any other orbital element, and since $\Omega$ itself seems to be of 
little practical importance.

In terms of  the hard-binary separation $a_h$ (Eq.~\ref{Equation:a_h}) 
and the initial hardening timescale (Eq.~\ref{Equation:t_h}),
we can write Eqs.~(\ref{Equation:Rate1}) as
\bsub
\label{Equation:Rate2}
\barr
\frac{d(a/a_h)}{d(t/t_h)} &=& - \left(\frac{a}{a_h}\right)^2 \left(\frac{S}{S_h}\right) - \left(\frac{a_{\rm GR,0}}{a_h}\right)^5 \left(\frac{a}{a_h}\right)^{-3}\; F(e) , \label{Equation:dadt} \\
\frac{de}{d(t/t_h)} &=&  \left(\frac{a}{a_h}\right) \left(\frac{S}{S_h}\right) K - \frac{19}{12} \left(\frac{a_{\rm GR,0}}{a_h}\right)^5 \left(\frac{a}{a_h}\right)^{-4}\; G(e), \\
\frac{d\theta}{d(t/t_h)} &=& \left(\frac{a}{a_h}\right) \left(\frac{S}{S_h}\right) D, \\
a_{\rm GR,0} &\equiv& \left(\frac{64}{5}\frac{\mathcal{Q} G^3M_{12}^3}{c^5 S_h}\right)^{1/5}.
\label{Equation:a_GR}
\earr
\esub
The physical meaning of $a_{\rm GR,0} $ is the binary separation at which the hardening rate due to GW emission
 equals to that due to stellar interactions for a circular-orbit binary assuming $S=S_h$ (cf. Fig.~\ref{Figure:e,theta(a)}). 
 
In the case of an infinite homogenous distribution of stars with density $\rho$ and velocity dispersion $\sigma$,
\barr\label{Equation:S_h}
S_h = H\frac{G\rho}{\sigma},
\earr
where $H\approx15$ has been determined by scattering experiments \citep{Quinlan1996}. 
In a real (inhomogeneous) galaxy, $\rho$ and $\sigma$ in Eq.~(\ref{Equation:S_h}) should be the density and velocity dispersion at (approximately) the influence radius of the binary, $r_{\rm infl}$ (as shown in \cite{PaperI}). 
Eqs. (\ref{Equation:a_GR}) and (\ref{Equation:S_h}) combine to give
\barr\label{Equation:a_GR(rho)}
\frac{a_{\rm GR,0}}{a_h} &\approx& 4.9\times 10^{-3} \mathcal{Q}^{-4/5} 
\left(\frac{M_{12}}{10^8 \msun}\right)^{1/25} 
\left(\frac{\rho}{10^3\msun\mathrm{pc}^{-3}}\right)^{-1/5} .
\earr
Here we have used the $M-\sigma$ relation \cite{msigma}:
\barr\label{Equation:M-sigma}
\frac{M_{12}}{10^8 \msun} \approx 1.66 \left(\frac{\sigma}{200\,\kms}\right)^5.
\earr
Since $\rho\sim M_{12}/r_{\rm infl}^3$ and $\sigma^2\sim GM_{12}/r_{\rm infl}$, 
\barr\label{Equation:S_h(r_infl)}
S_h = S_\mathrm{infl} = b\sqrt\frac{GM_{12}}{r_{\rm infl}^5},
\earr
where $b=3\dots5$ depending on galaxy structure \citep{Vasiliev2015}; in what follows we set $b=4$.
We can therefore write another expression for $a_{\rm GR,0}$:
\barr\label{Equation:a_GR(r_infl)}
\frac{a_{\rm GR,0}}{a_h} &\approx& 2.6\times 10^{-3} \mathcal{Q}^{-4/5} 
\left(\frac{M_{12}}{10^8 \msun}\right)^{-1/10} 
\left(\frac{r_{\rm infl}}{10\mathrm{\,pc}}\right)^{1/2} .
\earr
To further simplify this expression, we can assume $r_{\rm infl} = GM_{12}/\sigma^2$ with $\sigma$ 
related to $M_{12}$ through Eq. (\ref{Equation:M-sigma}), which yields
\barr\label{Equation:r_infl}
r_{\rm infl} \approx 13.2\,\mathrm{pc}\left(\frac{M_{12}}{10^8 \msun}\right)^{3/5},
\earr
and substituting (\ref{Equation:r_infl}) into (\ref{Equation:a_GR(r_infl)}),
\barr\label{Equation:a_GR(M_12)}
\frac{a_{\rm GR,0}}{a_h} &\approx& 3.0\times 10^{-3} \mathcal{Q}^{-4/5} 
\left(\frac{M_{12}}{10^8 \msun}\right)^{1/5}  .
\earr
Henceforth we define $a_{\rm GR,0}$ via Eq.~(\ref{Equation:a_GR(M_12)}). 

Since $a$ decreases monotonically with time, we can adopt $x\equiv a_h/a$ as a new time variable ($x$ increases with time), 
starting our simulations at $x=1$.
The evolution equations for $e$ and $\theta$ become
\bsub
\label{Equation:Rate3}
\barr
\frac{de}{dx} &=& \left[
K - \frac{19}{12} \left(\frac{S_h}{S}\right) \left(\frac{a_{\rm GR,0}}{a_h}\right)^5 x^5 G(e)
\right] \left[x + \left(\frac{S_h}{S}\right) \left(\frac{a_{\rm GR,0}}{a_h}\right)^5 x^6 F(e)\right]^{-1}, \\
\frac{d\theta}{dx} &=& D \left[x + \left(\frac{S_h}{S}\right) \left(\frac{a_{\rm GR,0}}{a_h}\right)^5 x^6 F(e)\right]^{-1}.
\earr
\esub

We adopt the following expressions from \citep{PaperI} for $K$ and $D$:
\bsub
\label{Equation:K,D}
\barr
K &=& 1.5\,e\, (1-e^2)^{0.7}\, [0.15-(2\eta-1)\cos\theta],\\
D &=& -0.3\,(2\eta-1) \sqrt\frac{1+e}{1-e} \sin\theta .
\earr
\esub
These expressions have been calculated assuming a hard binary ($a<a_h$),
hence they do not depend on $a$ or $t$. 
The hard-binary assumption is justified here because we choose $a=a_h$ as the initial separation.

\begin{figure}[h!]
\includegraphics[width=0.49\textwidth]{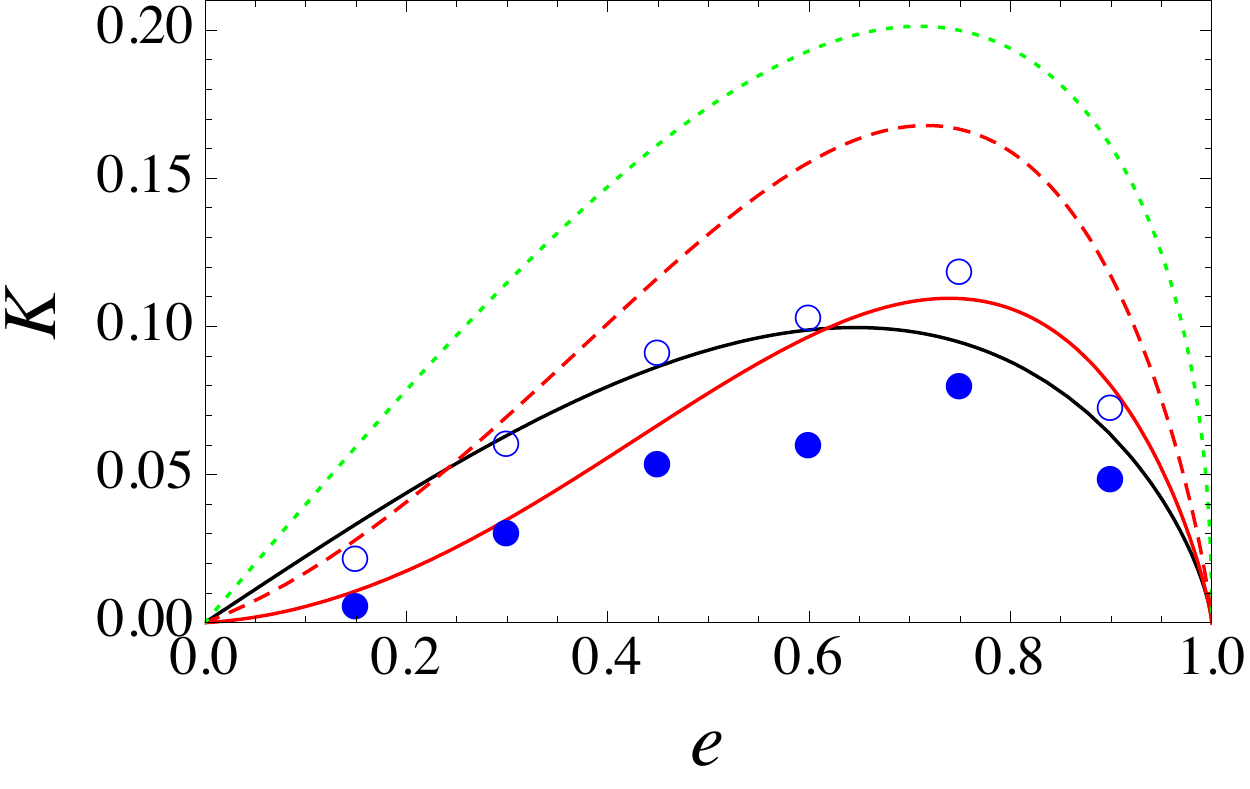}
\includegraphics[width=0.49\textwidth]{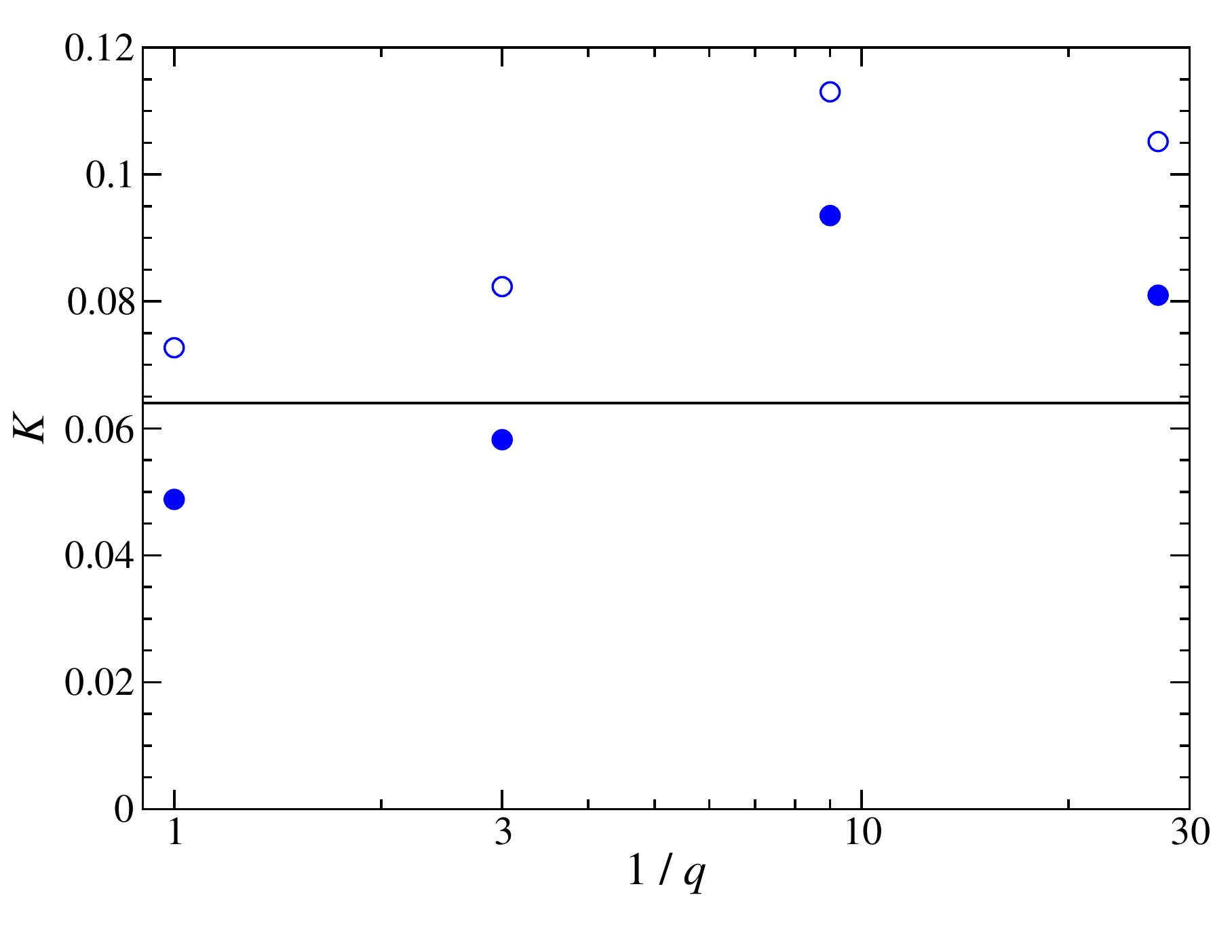}
\caption{Dimensionless eccentricity growth rate $K$ (Eq.~\ref{Equation:K,D}a)
for equal-mass binaries and varying eccentricity (left) or $e=0.9$ and varying mass ratio (right) in nonrotating nuclei. 
Black curve is our expression (\ref{Equation:K0}). 
Green dotted curve: the expression of \citet{Mikkola1992} in $a\rightarrow0$ limit. 
Red curves: the results of \citet{Quinlan1996} for $a/a_h=0.16$ (solid) and $a/a_h=0.018$ (dashed). 
Blue circles: the results of \citet{Sesana2006} for $a/a_h=0.16$ (filled) and $a/a_h=0.018$ (empty).}
\label{Figure:K(e)}
\end{figure}

In an irrotational nucleus ($\eta=1/2$), the rate coefficients (\ref{Equation:K,D}) become
\barr\label{Equation:K0}
K = 0.225\,e\, (1-e^2)^{0.7},\ \ \ \ D = 0.
\earr
A few previous papers \citep{Mikkola1992, Quinlan1996, Sesana2006} have calculated $K$ for a non-rotating nucleus; as shown in Fig.~\ref{Figure:K(e)}, our expressions are consistent with those of \citet{Sesana2006} in the $a/a_h\rightarrow0$ limit.

Fig. \ref{Figure:e,theta(a)} presents solutions to the coupled Eqs.~(\ref{Equation:Rate3}), 
for two different degrees of nuclear rotation, $\eta=0.6$ (low rotation) and $\eta=1$ (maximal rotation). 
The binary's orbital inclination always decreases, so that initially counterrotating binaries ($\theta\approx\pi$) tend to become corotating ($\theta\approx 0$). 
For corotating binaries, $e$ almost always decreases with time: first due to stellar encounters, and later due to GW emission.
In the case of counterrotating binaries, $e$ generally increases at early times but eventually starts to decrease -- either because
the binary has become corotating, or because of GW emission. 
For binaries in maximally-rotating galaxies, reorientation of the orbital plane takes place quickly, and as a result, 
the binary enters the GW-dominated regime with low eccentricity. 
On the other hand, in slowly-rotating galaxies, both reorientation and circularization are much less pronounced,
allowing the binary to enter the GW-dominated regime with high eccentricity. 

\begin{figure*}[h!]
	\centering
	\subfigure{\includegraphics[width=0.49\textwidth]{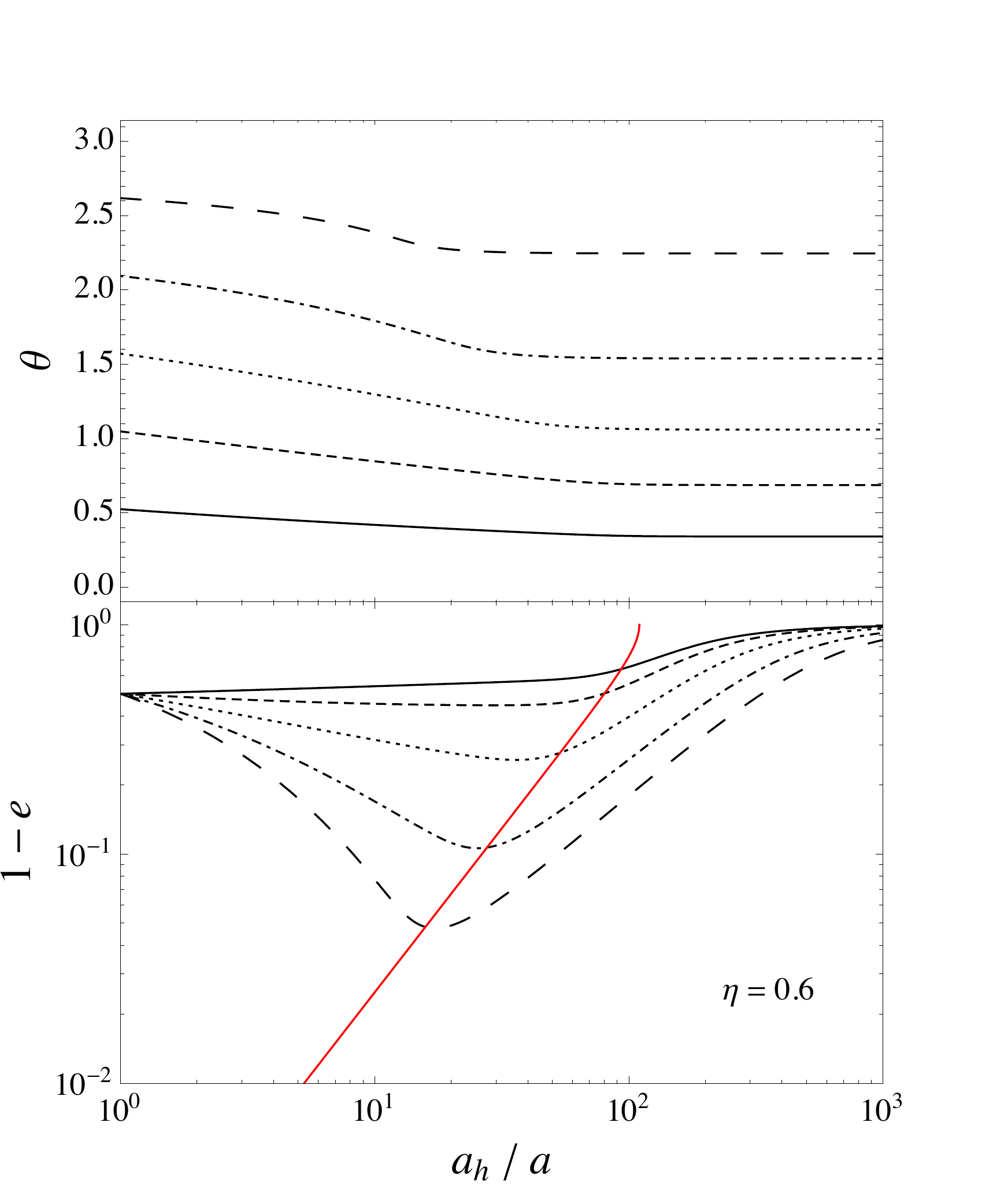}}
	\subfigure{\includegraphics[width=0.49\textwidth]{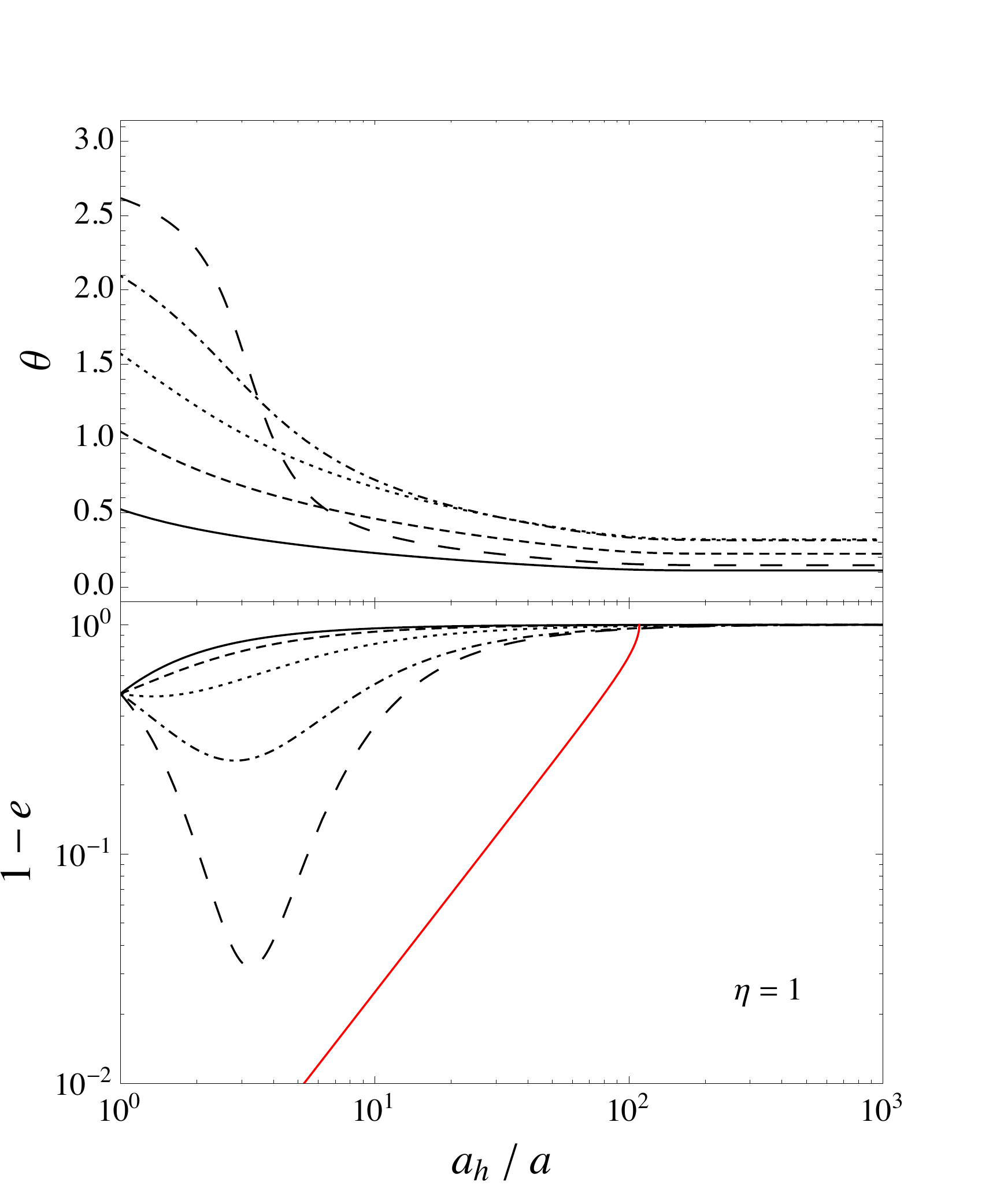}}
\caption{Evolution of orbital inclination $\theta$ and eccentricity $e$ for an equal-mass binary in 
a slowly-rotating nucleus ($\eta=0.6$, left) and (b) a maximally-rotating nucleus ($\eta=1$, right), 
computed using Eqs.~(\ref{Equation:Rate3}) and (\ref{Equation:K,D}) with $M_{12}=10^8\msun$ and $S=S_h$. 
Different line styles correspond to different initial values of $\theta$. 
The initial eccentricity is always $e_0=0.5$. 
The red curve separates the regimes where the hardening of the binary is dominated by stellar encounters (to the left) and GW emission (to the right); its equation is $a(e) = a_{\rm GR,0} F^{1/5}(e)$ (see Eq.~\ref{Equation:dadt}).}
\label{Figure:e,theta(a)}
\end{figure*}

By interacting with stars, a massive binary tends to decrease the number of stars on orbits that can interact with it 
(``loss-cone depletion'').
This depletion is accounted for by letting $S$ depend on $a$ or $t$; the ``full-loss-cone'' approximation 
corresponds to $S=\mathrm{const}$. 
Here we assume that loss-cone depletion has no effect on the mean energy or angular momentum carried away by a single stellar interaction; such an assumption is justified considering the chaotic nature of a binary-star interaction when the final velocity and orbital momentum of a star are weakly correlated with the initial ones. 
Then the change in hardening rate $S$ is due only to the change in the rate of stellar interactions with the binary: $S\propto dn/dt$. 
Since all of the orbital parameters change at rates that are proportional to $dn/dt$, they are all proportional to $S$ as well. 
That is why we can allow $S$ to depend on $a$ (or $t$) while keeping $K$ and $D$ time-independent. 

The rate of loss-cone depletion depends strongly on the ``geometry'', i.e. the shape, of the galaxy.
We adopt the following expressions for $S(a)$ from \citet{Vasiliev2015}:
\bsub\label{Equation:galaxyType}
\barr
S(a) &=& k S_\mathrm{infl} \left(\frac{a}{a_h}\right)^\alpha,\\
k&=&0.4, \;\;\alpha = 0.3 \;\; {\rm for\;triaxial\;nuclei},\\
k&=&(N_\star/10^5)^{-1/2}, \;\;\alpha = 0 \;\; {\rm for\;axisymmetric\;nuclei},\\
k&=&(N_\star/10^5)^{-1}, \;\;\alpha = 0 \;\; {\rm for\;spherical\;nuclei}
\earr
\esub
where $N_\star=M_\mathrm{gal}/M_\odot$ is the number of stars in the galaxy.
The $N_\star$-dependence in Eqs.~(\ref{Equation:galaxyType}c,d) reflects the fact that in 
spherical and axisymmetric geometries, conservation of angular momentum (spherical symmetry) 
or its component along the symmetry axis (axisymmetry)
fixes the minimum periapsis distance accessible to a star.
Once all the stars on an orbit with given periapsis have been removed, 
continued supply of stars to the binary is only possible after new stars have been scattered onto the orbit
by gravitational encounters, at rates that are $N_\star-$dependent. 
In triaxial galaxies, much of the phase space corresponds to orbits with no minimum periapsis;
the time for a star on such an orbit to reach the binary depends much more on torques from the
large-scale mass distribution than on two-body relaxation, hence the lack of an appreciable $N_\star$
dependence in the expression for the ``triaxial'' hardening rate.
We anticipate the discussion in the next section by mentioning that mergers between luminous,
gas-poor galaxies are expected to result in triaxial merger remnants, hence in efficient hardening of the binary.

Throughout this paper, we ignore the effect of torques from any ambient gas on the
evolution of the binary \sbh\ \cite{Ivanov1999}.
Gas is expected to be present, in dynamically significant densities, in low-mass
systems (bulges of disk galaxies; dwarf elliptical galaxies), and more generally
in galaxies at high redshift.
One justification for our neglect of gas-dynamical torques is the recent realization,
embodied here in Eqs. (\ref{Equation:galaxyType}), that stellar-dynamical interactions can be much
more effective than had previously been thought at evolving binary \sbhs\ to separations $\ll a_h$.
Nevertheless, there is a body of work, as summarized by \citep{Sesana2013b},
 that argues that gaseous torques could shorten the time spent by a massive binary in the later stages of evolution, when GW emission competes with stellar-dynamical interactions.

\section{Results}
\label{Section:Results}

\begin{figure*}[h!]
	\centering
	\subfigure{\includegraphics[width=0.49\textwidth]{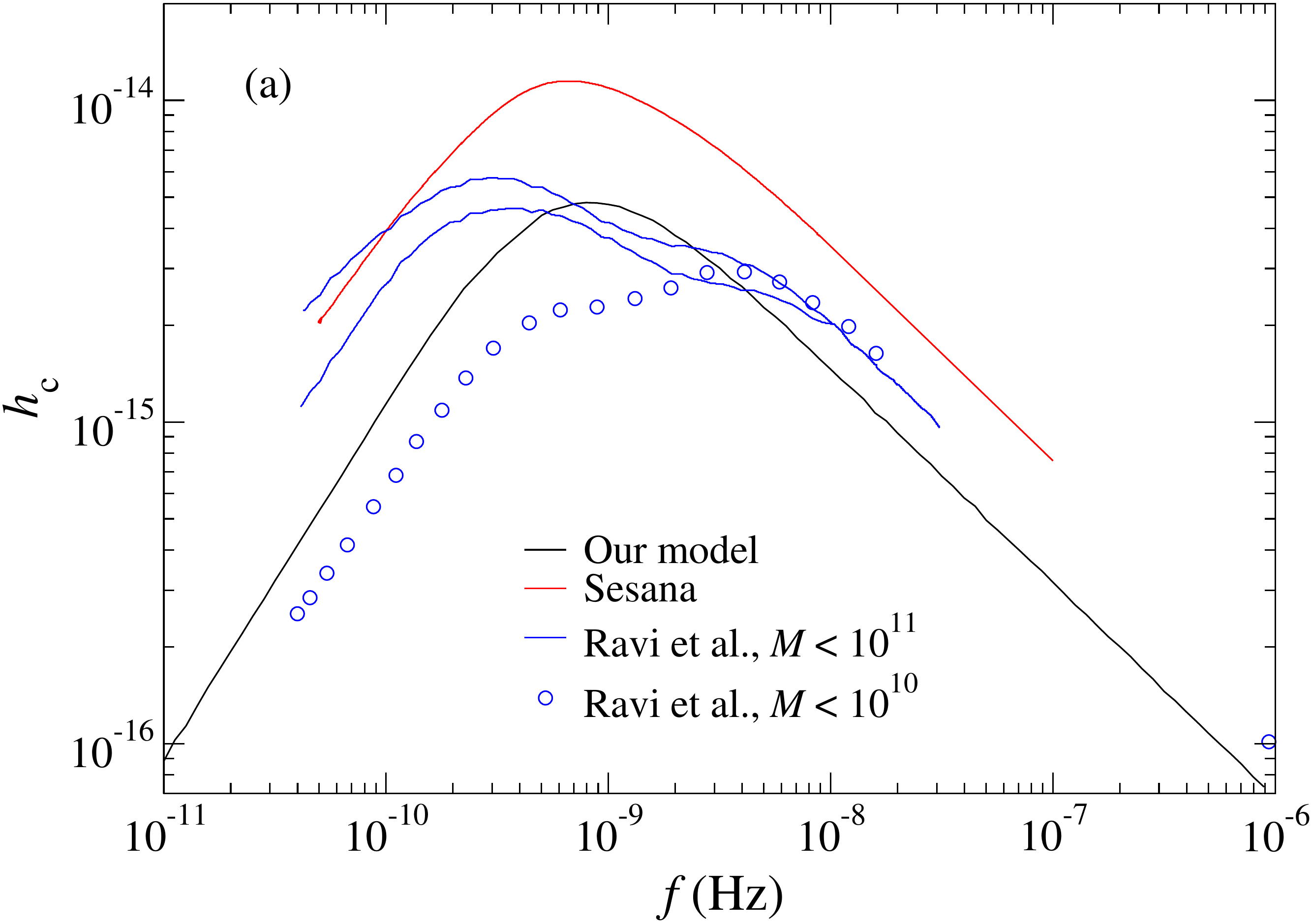}}
	\subfigure{\includegraphics[width=0.49\textwidth]{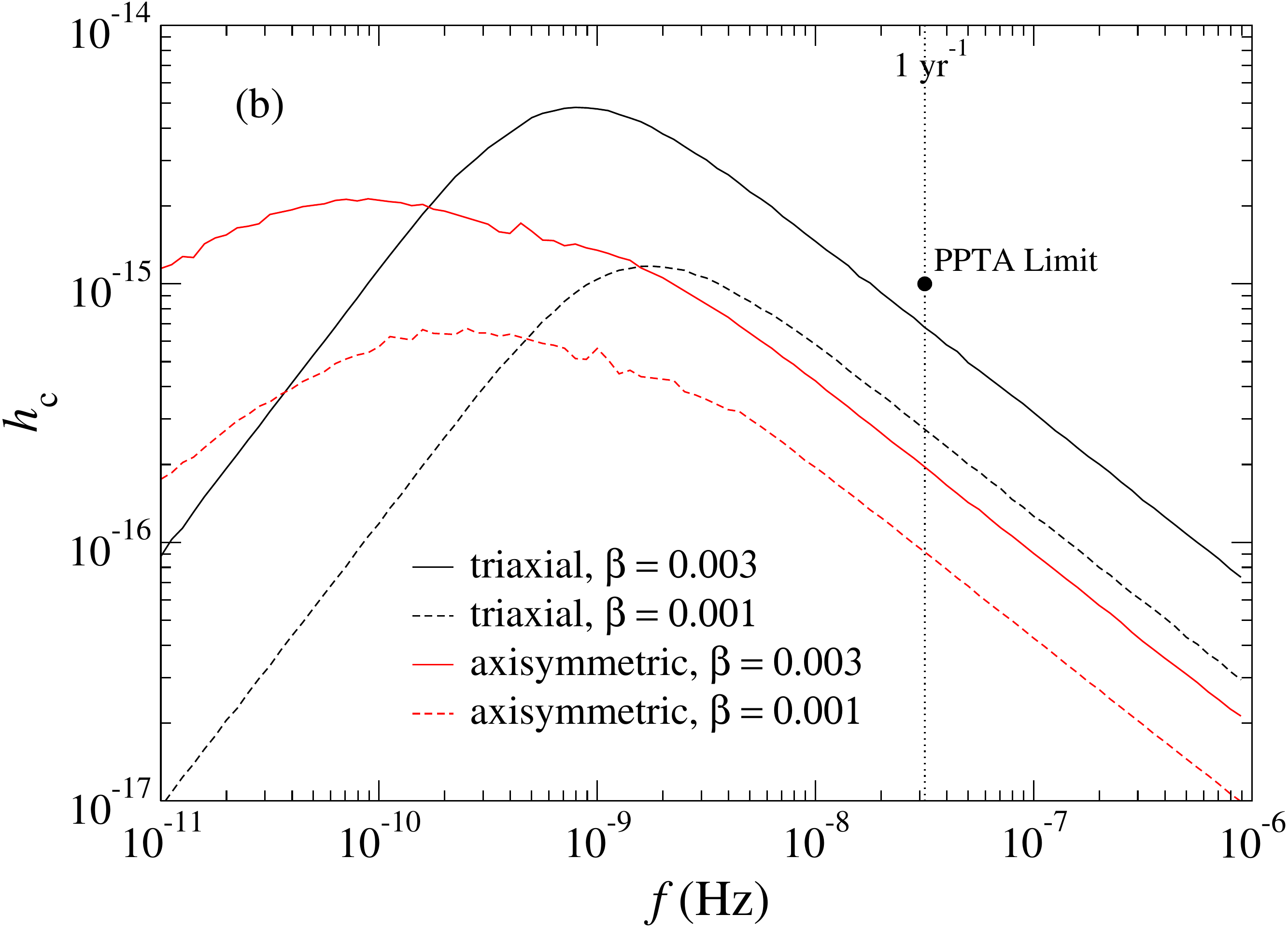}}
\caption{
(a) Predicted GW strain for circular binaries. Black: our model described in \S~\ref{Section:Method} assuming 
the ``triaxial'' (efficient) hardening law and $\beta=0.003$. 
Red: model from \citet{Sesana2013b}. Blue: models from \citet{Ravi2014}; solid lines correspond to $M_{12}/\msun=10^{6.5}...10^{11}$ and two different assumption about the stellar density profile; circles correspond to $M_{12}/\msun=10^{6.5}...10^{10}$. 
(b) Predicted GW strain for different assumptions about $\beta$ (\sbh\ mass) and binary hardening law. 
Black dot indicates the 95\%-confidence upper limit from PPTA \citep{Shannon2015}. 
}
\label{Figure:e=0}
\end{figure*}

We first consider the case in which all binary orbits are initially circular (and remain so).
This assumption leaves only two important parameters: galaxy geometry 
(that is, the binary hardening law of Eq.~\ref{Equation:galaxyType})
and $\beta$ (the ratio of \sbh\ mass to bulge mass). 
In Fig. \ref{Figure:e=0}a we plot $h_c(f)$ as predicted by our model after setting $\beta=0.003$ and assuming 
``triaxial'' (i.e. efficient) binary hardening.
Also plotted there are the results of \citet{Sesana2013b} and \citet{Ravi2014}, who made similar assumptions about $\beta$
and binary hardening rates.
Given the uncertainties quoted by those authors -- 95\% confidence intervals in $h_c$ are said to be $\pm0.5\mathrm{\,dex}$  \citep{Sesana2013a} -- we conclude that our model is consistent with both of them. 

\begin{figure*}[h!]
	\centering
	\subfigure{\includegraphics[width=0.49\textwidth]{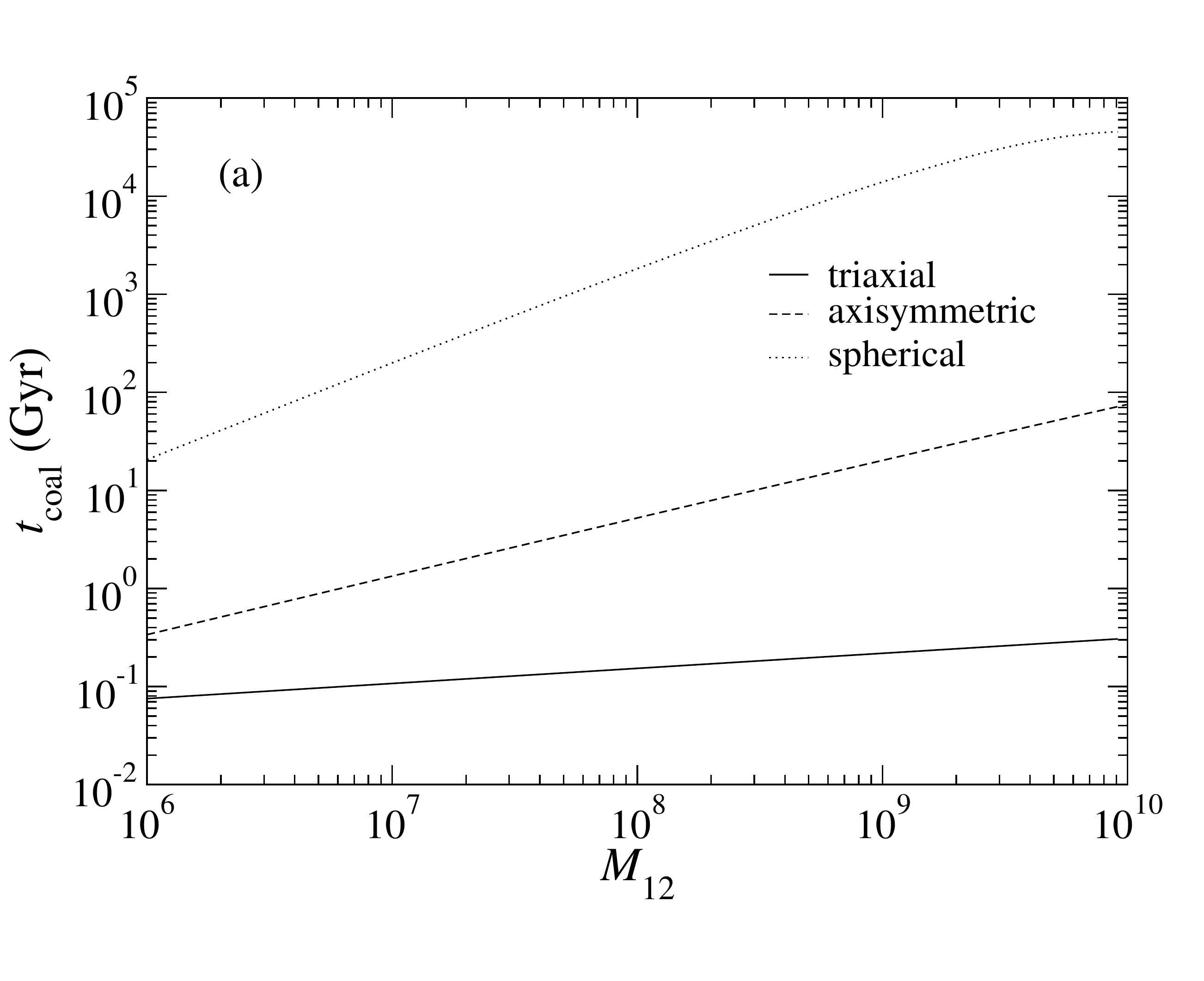}}
	\subfigure{\includegraphics[width=0.49\textwidth]{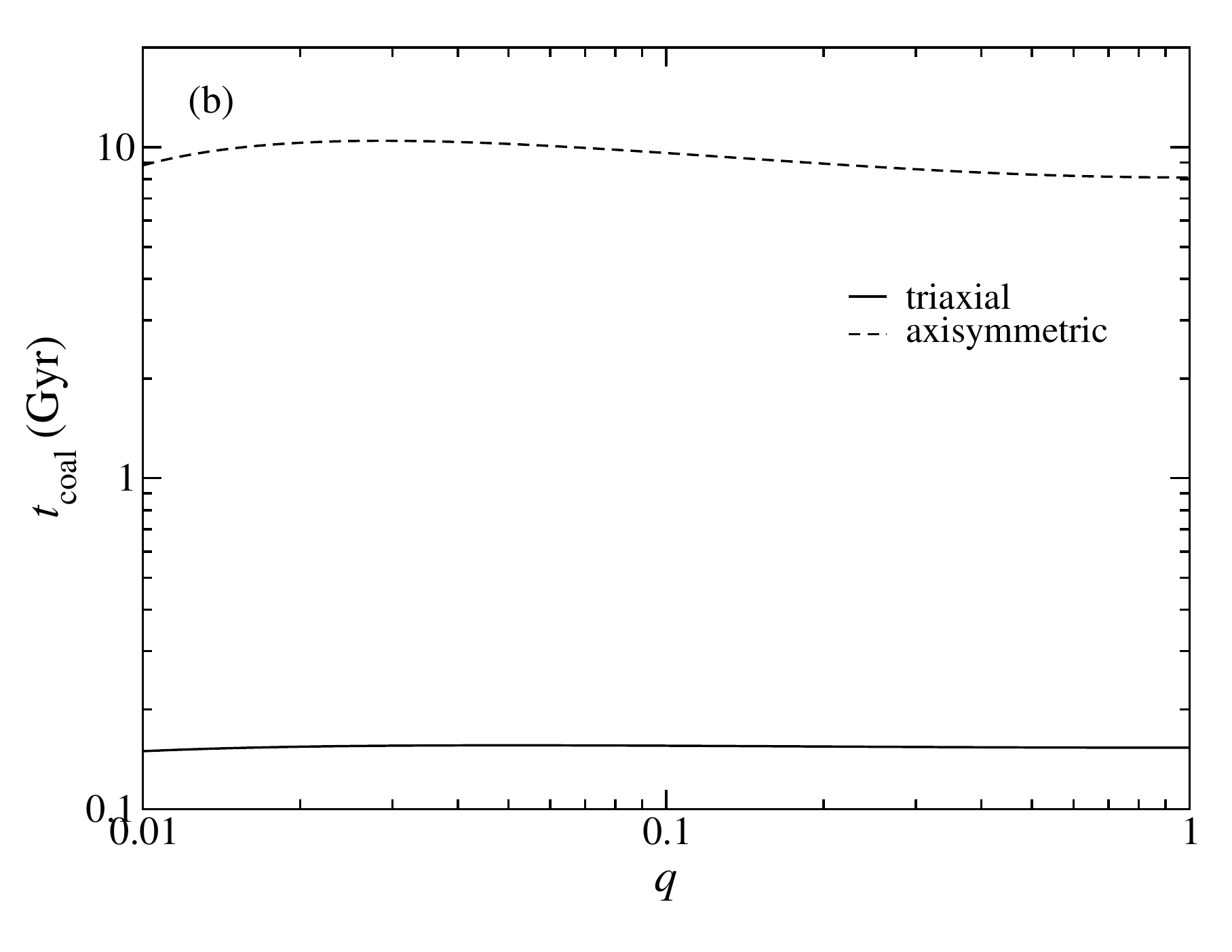}}
	\subfigure{\includegraphics[width=0.49\textwidth]{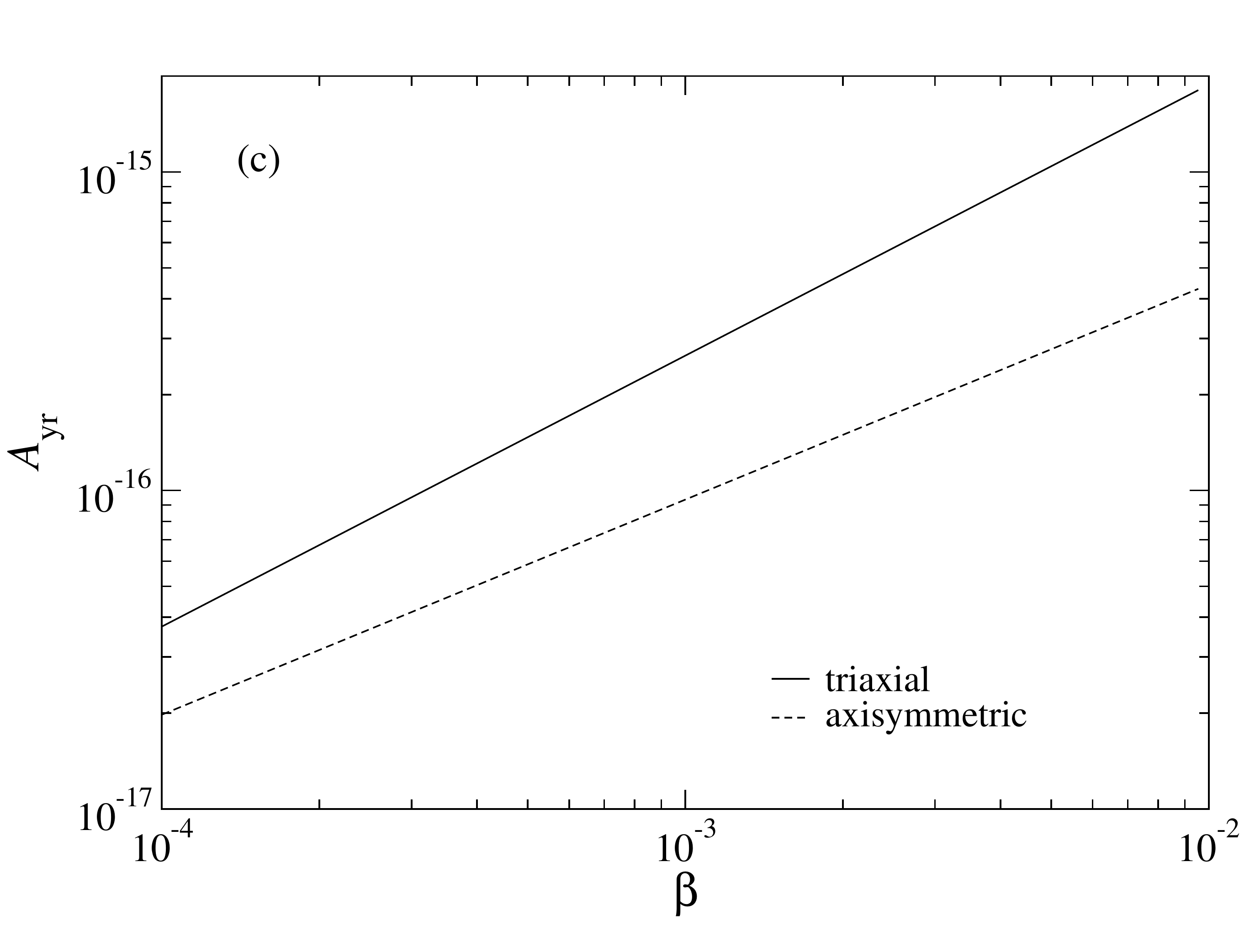}}
\caption{
(a) Coalescence time (from $a=a_h$ to $a\approx0$) as a function of $M_{12}$ for equal-mass circular binaries 
in triaxial, axisymmetric and spherical galaxies. (b) Coalescence time as a function of $q$ for $M_{12} = 10^8\msun$.
(c) Strain amplitude (Eq.~\ref{Equation:A_yr}) for circular-orbit 
binaries in triaxial and axisymmetric galaxies as a function of $\beta$ (for spherical galaxies $A_{\rm yr}=0$).
}
\label{Figure:t,A,e=0}
\end{figure*}

\begin{figure}[h!]
	\centering
	\subfigure{\includegraphics[width=0.49\textwidth]{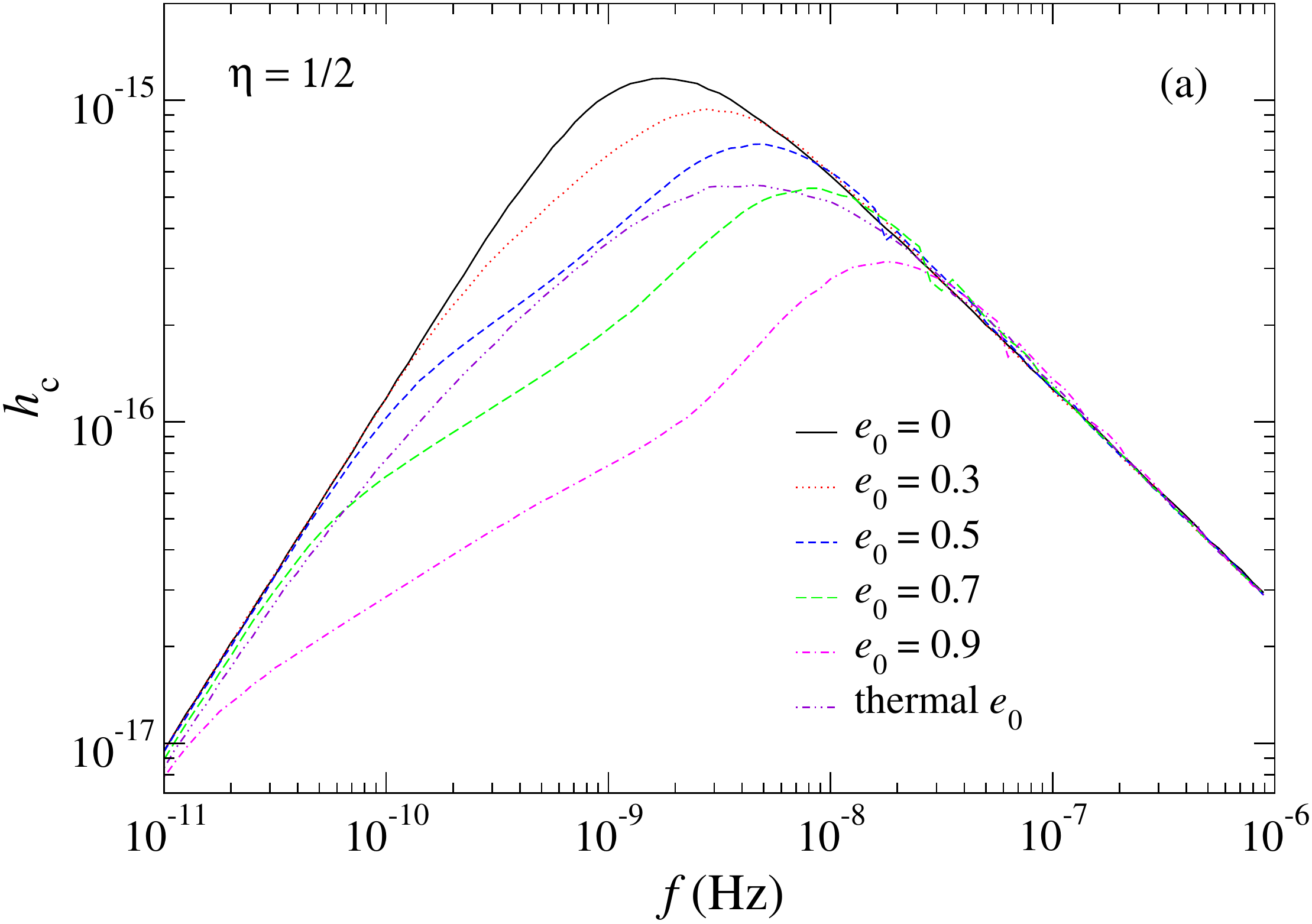}}
	\subfigure{\includegraphics[width=0.49\textwidth]{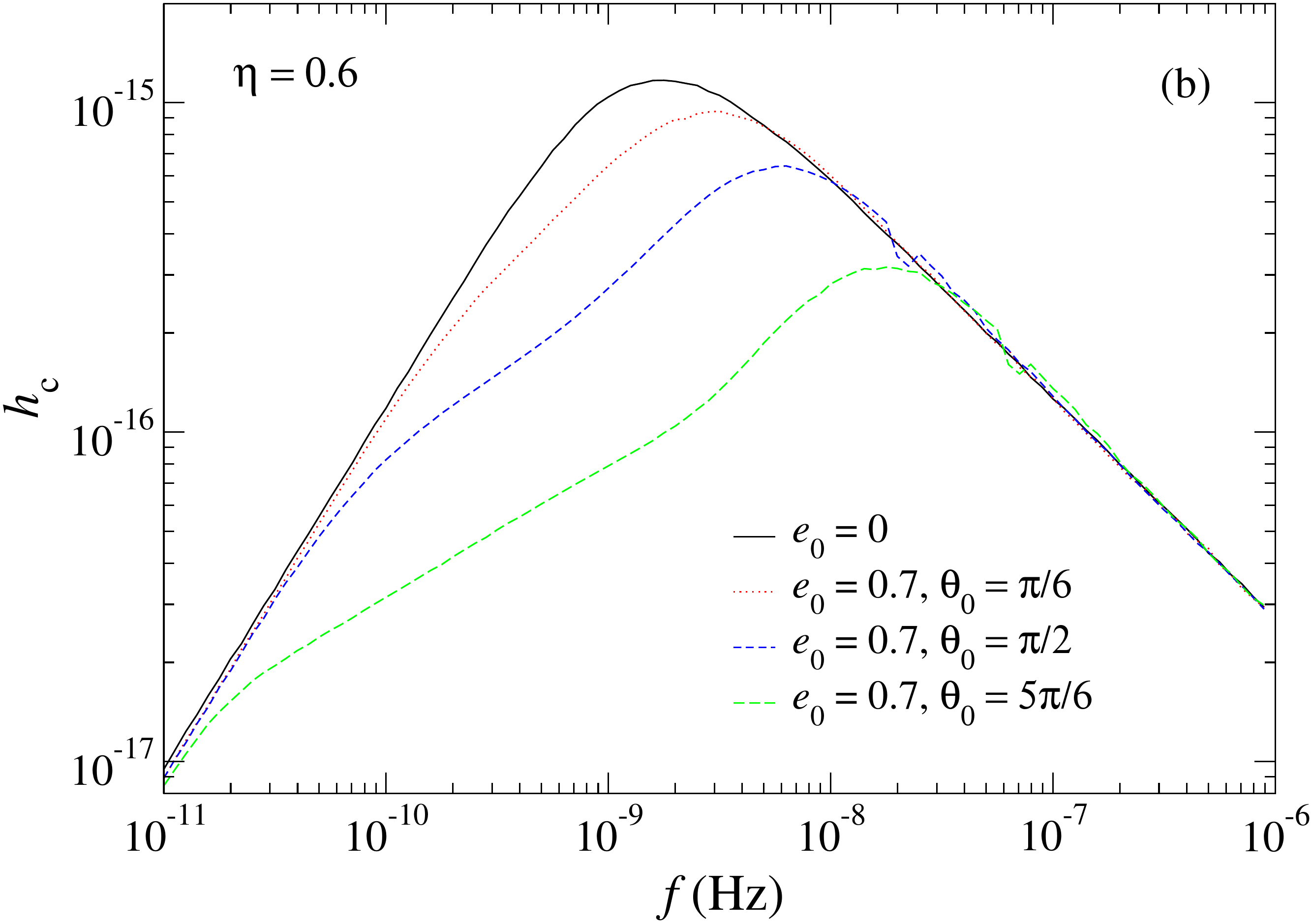}}
	\subfigure{\includegraphics[width=0.49\textwidth]{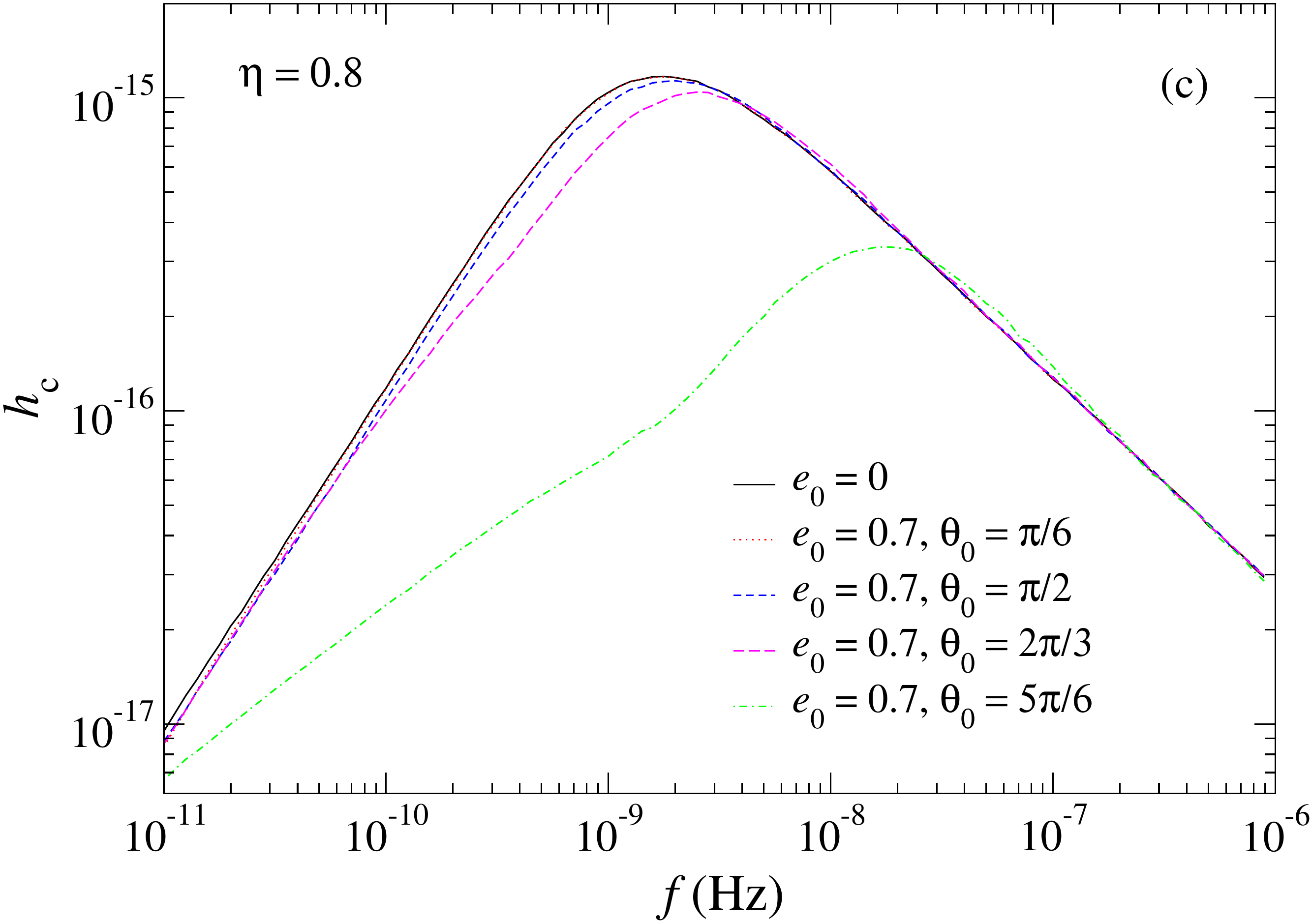}}
	\subfigure{\includegraphics[width=0.49\textwidth]{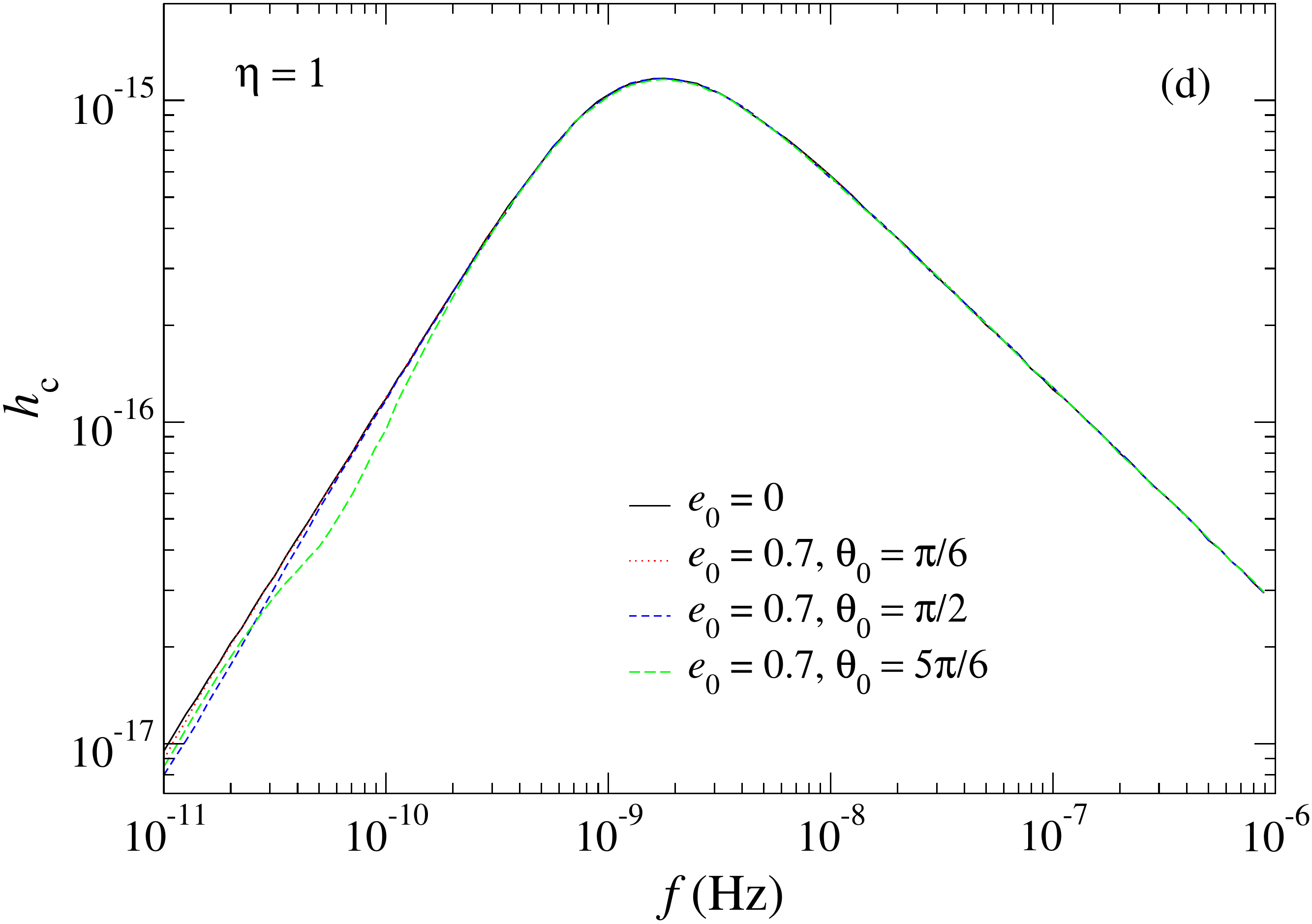}}
\caption{
Predicted GW strain for four different values of the nuclear corotation fraction $\eta$.
All curves assume $\beta=0.001$. Initial orbital elements $(e_0,\theta_0)$ are assumed to be the same for all binaries.
}
\label{Figure:e>0}
\end{figure}

Fig. \ref{Figure:e=0}b illustrates the dependence of $h_c(f)$ on galaxy morphology (i.e. binary hardening law) and $\beta$. 
Decreasing the assumed \sbh\ masses reduces the GW emission at all frequencies and shifts the peak of the spectrum to higher frequencies,
since less-massive binaries enter the GW-dominated regime at higher orbital frequencies, i.e., smaller semimajor axes (Eq.~\ref{Equation:a_GR(M_12)}). 
Changing the assumed galaxy morphology from triaxial to axisymmetric implies significant reduction in binary hardening rates
 (Eq.~\ref{Equation:galaxyType}).
 There are two consequences.
As shown in Fig.~\ref{Figure:t,A,e=0}a-b, binaries in axisymmetric galaxies with $M_{12}\gtrsim4\times10^8\msun$ and any $q$
have coalescence times that are longer than a Hubble time; at present they might not have reached the GW-dominated regime.
This results in $h_c$ being $\sim3$ times lower at high frequencies compared with the ``triaxial'' case. 
At the same time, $h_c$ in the ``axisymmetric'' case is higher at low frequencies because the binaries
spend more time radiating at large orbital separations. 
In the case of spherical galaxies (not shown here), coalescence times are so long 
that there is essentially no GW emission at PTA-accessible frequencies. 

The coalescence timescales in our nonrotating models are $6-7$ times shorter than those found by \citet{Vasiliev2015}. 
This difference is a consequence of different definitions of the influence radius: 
we define it as $GM/\sigma^2$ while Vasiliev et al. use an empirical relation between the observed $r_{\rm infl}$ and black hole mass, obtained via the $M-\sigma$ relation (Eq. 9 in their paper), which implies larger values of $r_{\rm infl}$. 

At sufficiently high frequencies, where binary dynamics are dominated by GW emission, 
the characteristic strain has the power-law dependence of Eq.~(\ref{Equation:hcPhinney}):
\barr\label{Equation:A_yr}
h_c(f) = \ayr \left(\frac{f}{1\,\mathrm{yr}^{-1}}\right)^{-2/3}. \nonumber
\earr
Fig. \ref{Figure:t,A,e=0}b shows the dependence of $\ayr$ on $\beta$ assuming circular orbits ($e=0$):
\bsub\label{Equation:A_yr(beta)}
\barr
\ayr &\approx& 2.7\times10^{-16}\, \left(\frac{\beta}{10^{-3}}\right)^{0.85}\;\; {\rm for\;triaxial\;galaxies},\\
\ayr &\approx& 9.3\times10^{-17}\, \left(\frac{\beta}{10^{-3}}\right)^{0.68}\;\; {\rm for\;axisymmetric\; galaxies}. \label{Equation:A_yr(beta)-axisymmetric}
\earr
\esub
(Note that the degree of nuclear rotation, $\eta$, is unimportant in the circular-orbit case.)
Among all the possible parameter combinations, the choice ``$\beta=0.003$ + triaxial galaxies'' should yield results most similar to those in the recent studies of \citet{Sesana2013a}, \citet{Ravi2014}, and \citet{Simon2016}.
We indeed find that our estimate $\ayr\approx6.8\times10^{-16}$ is consistent, within $1\sigma$, with those in the aforementioned papers.


We have found the following formula to be a good analytical approximation for the ``$e=0$ + triaxial galaxies'' case (which is a reasonable assumption, as we'll show in the end of this section):

\bsub\label{Equation:A,f_b}
\barr
h_c(f) &=& A\frac{(f/f_{\rm yr})^{-2/3}}{1+(f_{\rm b}/f)^{53/30}},\\
A &=& 2.77\times10^{-16}\, \left(\frac{\beta}{10^{-3}}\right)^{0.83} \sqrt{\frac{\dot{\cal N}_m}{\dot{\cal N}_{m,0}}} ,\\
f_{\rm b} &=& 1.35\times10^{-9}\,{\rm Hz}\, \left(\frac{\beta}{10^{-3}}\right)^{-0.68}.
\earr
\esub

Here we have accounted for the possibility of the actual galaxy merger rate $\dot{\cal N}_m$ being different than the one given by Eq.~(\ref{Equation:Xu}), $\dot{\cal N}_{m,0}$; see Section~\ref{Section:mergerRate} for the discussion of the possible reasons for that. We assume that $\dot{\cal N}_m$ always has the same dependence on galaxy mass and redshift and only the scale factor may vary, which leaves the shape of the spectrum unchanged and only changes its amplitude (Eq.~\ref{Equation:hcfinal}).

This approximation is accurate to within $5\%$ at PTA-sensitive frequencies $f>10^{-9}\,\mathrm{Hz}$. It's similar to the one suggested by Sampson et al. (Eq. \ref{Equation:Sampsonhc}) for single-mass binary population, but has a slightly different low-frequency slope (1.1 vs. 1, because of the difference in stellar hardening rate) and a broader peak (due to different BH masses contributing to the signal). Note that since $f_{\rm b}<f_{\rm yr}$ at all reasonable values of $\beta$,

\barr
A = \ayr \left[1+\left(\frac{f_{\rm b}}{f_{\rm yr}}\right)^{53/30}\right] \approx \ayr
\earr

It is also true that $f_{\rm b}<0.1f_{\rm yr}$ (minimum frequency currently probed by PTA) unless $\beta\lesssim3\cdot10^{-4}$, which means it's unlikely that we will be able to see the break in stochastic GW background spectrum in the near future. However, this is true only in the assumption of zero eccentricity or quickly decreasing eccentricity, as we show below.


Fig. \ref{Figure:e>0} illustrates the impact of nonzero orbital eccentricities on the GW spectrum. It confirms the previous results that high eccentricity reduces GW emission at low frequencies \cite{Ravi2014,Huerta2015,Schnittman2015}.
As shown in Fig. \ref{Figure:e>0}a, for nonrotating galaxies, the higher the assumed initial eccentricity, 
the stronger this effect is. 
At high frequencies ($f\gtrsim10^{-8}\,\mathrm{Hz}$) the strain is unchanged because by the end of its dynamical evolution the binary orbit is always nearly circular due to GW emission. 
For rotating galaxies, the parameter $\theta_0$, the initial inclination of the binary's orbit, comes into play. 
The closer $\theta_0$ is to $\pi$ (= counterrotation), 
the greater the maximum eccentricity reached by the binary (cf. Fig. \ref{Figure:e,theta(a)}). 
However, as Figs. \ref{Figure:e>0}b-d show, the influence of $\theta_0$ is a strong function of a galaxy's degree of rotation. 
In maximally-rotating galaxies (Fig. \ref{Figure:e>0}d) the binary becomes corotating and eccentricities fall to negligible values before the binary starts to emit at PTA frequencies (Fig.~\ref{Figure:e,theta(a)}, right).
As a result, the spectrum is almost identical to that produced by circular binaries. 

Fig.~\ref{Figure:h_c-final} shows computed spectra assuming a ``thermal'' distribution of initial eccentricities and 
different combinations of the $\theta_0$- and $\eta$ distributions discussed previously (Eqs. \ref{Equation:e_0-thermal}, \ref{Equation:vsigmasesana} and \ref{Equation:F1(eta)}). 
As expected, for triaxial galaxies (Fig.~\ref{Figure:h_c-final}a), $h_c(f)$ is mildly (up to $\sim1.5$ times) attenuated compared with circular binaries at peak frequencies ($f\sim2\times10^{-10}\dots 4\times10^{-9}\,\mathrm{Hz}$) with almost no difference at other frequencies. For axisymmetric galaxies (Fig.~\ref{Figure:h_c-final}b) the effect of eccentricity is different: high initial eccentricities decrease the coalescence time, allowing more binaries to reach the GW-dominated stage and thus contribute to the GW background at high frequencies (up to $\sim2$ times increase in $h_c(f)$). 

Fig.~\ref{Figure:A_axi,e>0} shows in more detail the dependence of $\ayr$ on $(\eta,e_0,\theta_0)$. 
Comparison with the plots of $\tcoal$ illustrates the fact that the increase of $\ayr$ at high eccentricities is due to shorter 
coalescence timescales. 
In particular, $\ayr$ for axisymmetric and triaxial galaxies become comparable when $\tcoal$ is comparable, 
which can happen for high ($e_0\gtrsim0.9$) initial eccentricities. 
Also, as the upper-right panel shows, high values of $\eta$ usually imply smaller $\ayr$ values unless $\theta_0$ is high. 
$\ayr$ for triaxial galaxies is not shown on these plots because it is practically independent of eccentricity. 

\begin{figure}[h!]
	\centering
	\subfigure{\includegraphics[width=0.49\textwidth]{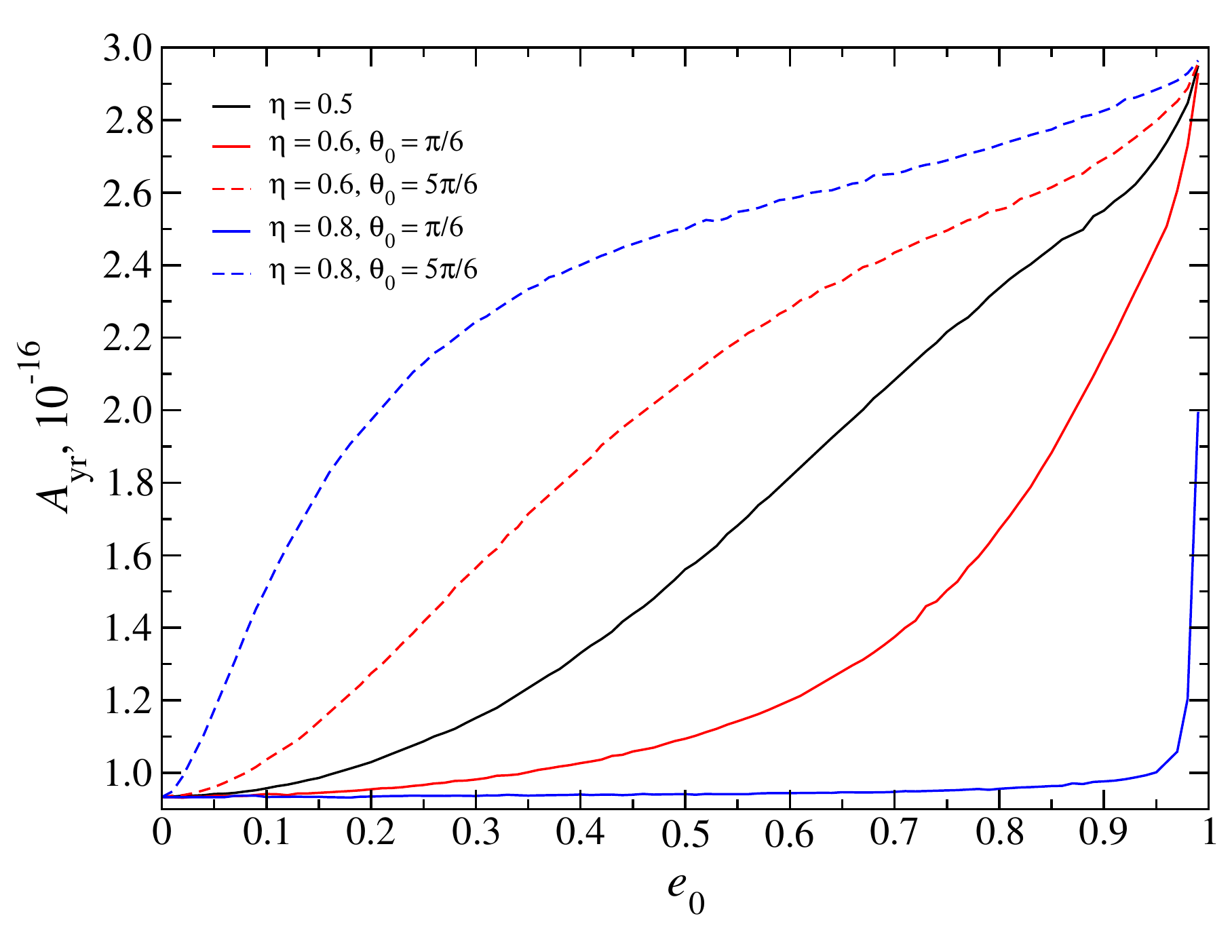}}
	\subfigure{\includegraphics[width=0.49\textwidth]{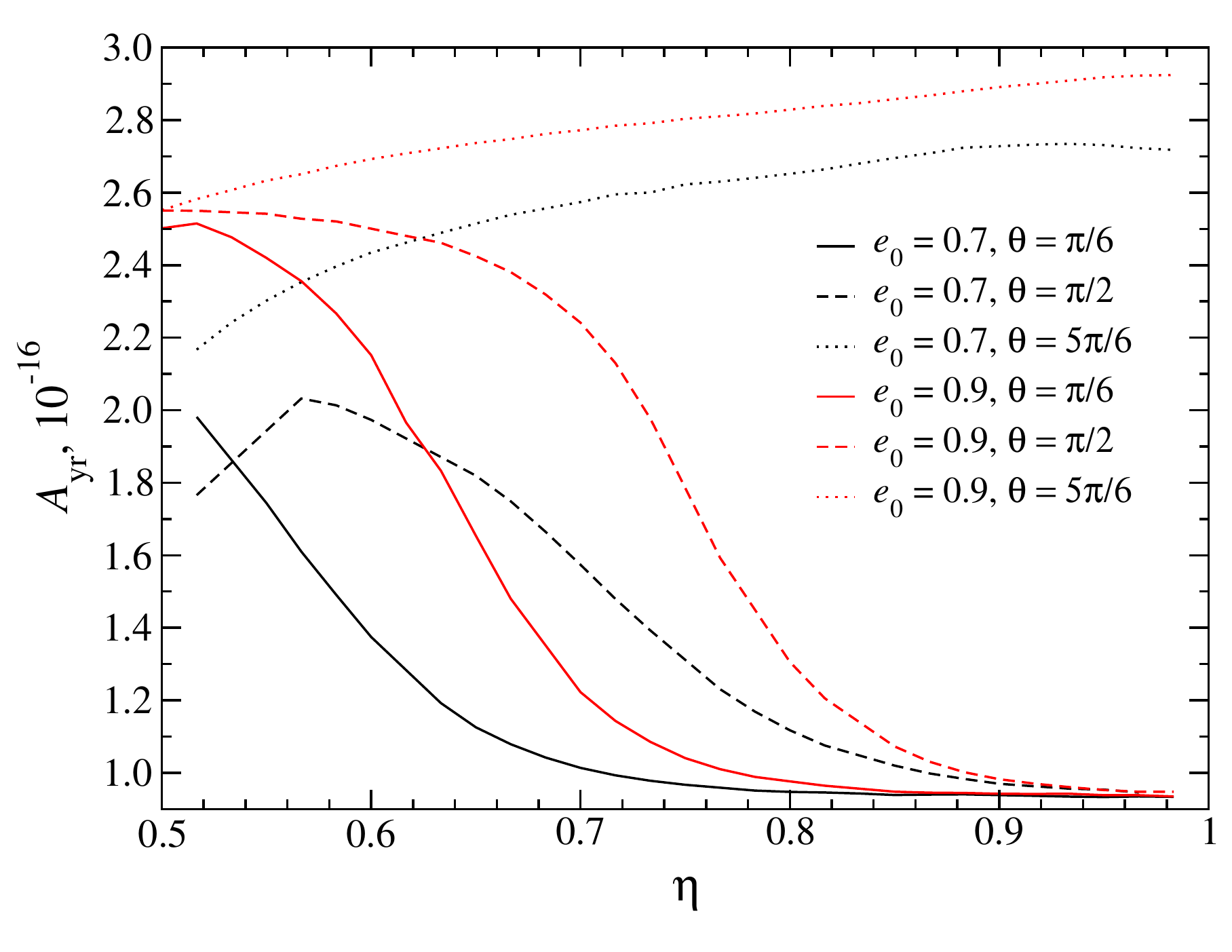}}
	\subfigure{\includegraphics[width=0.49\textwidth]{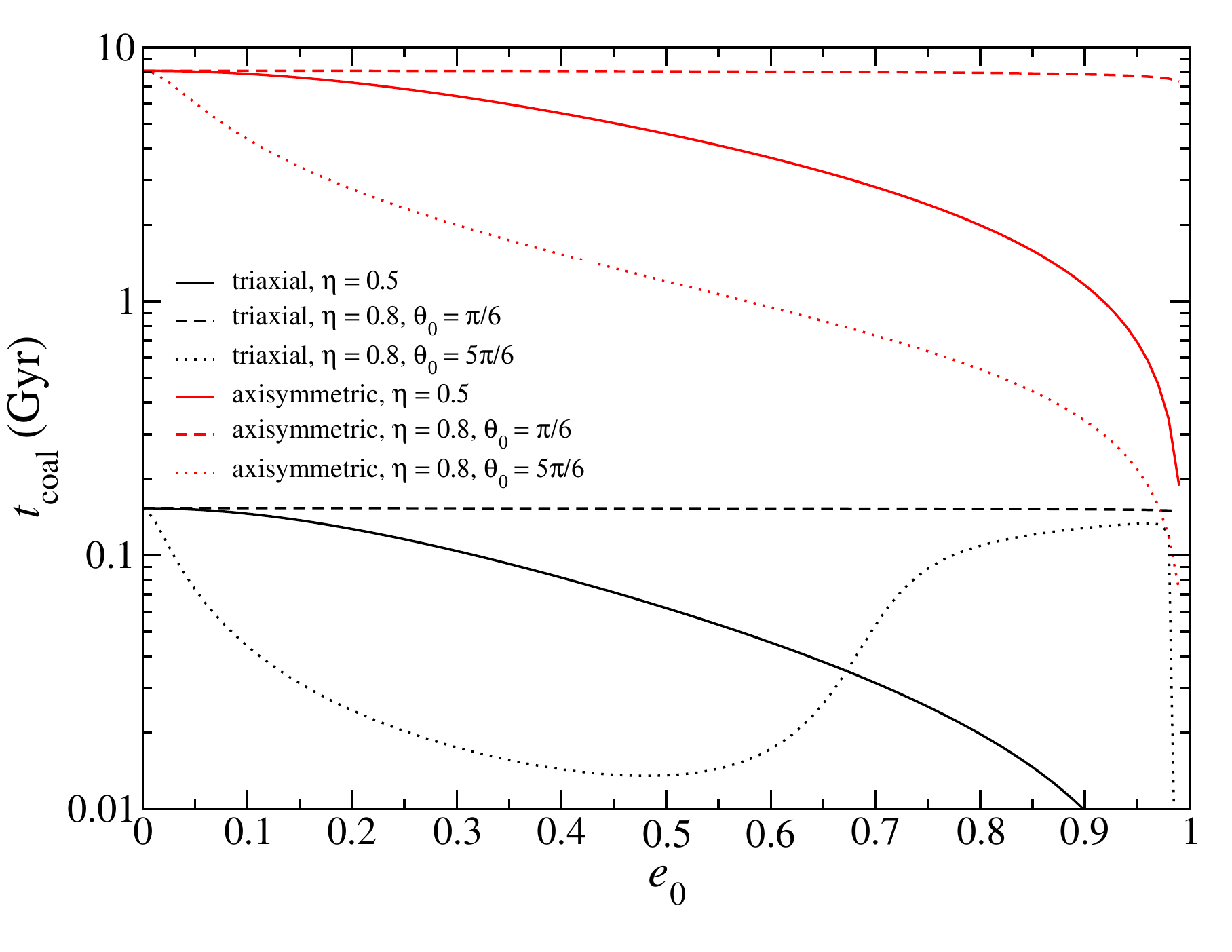}}
	\subfigure{\includegraphics[width=0.49\textwidth]{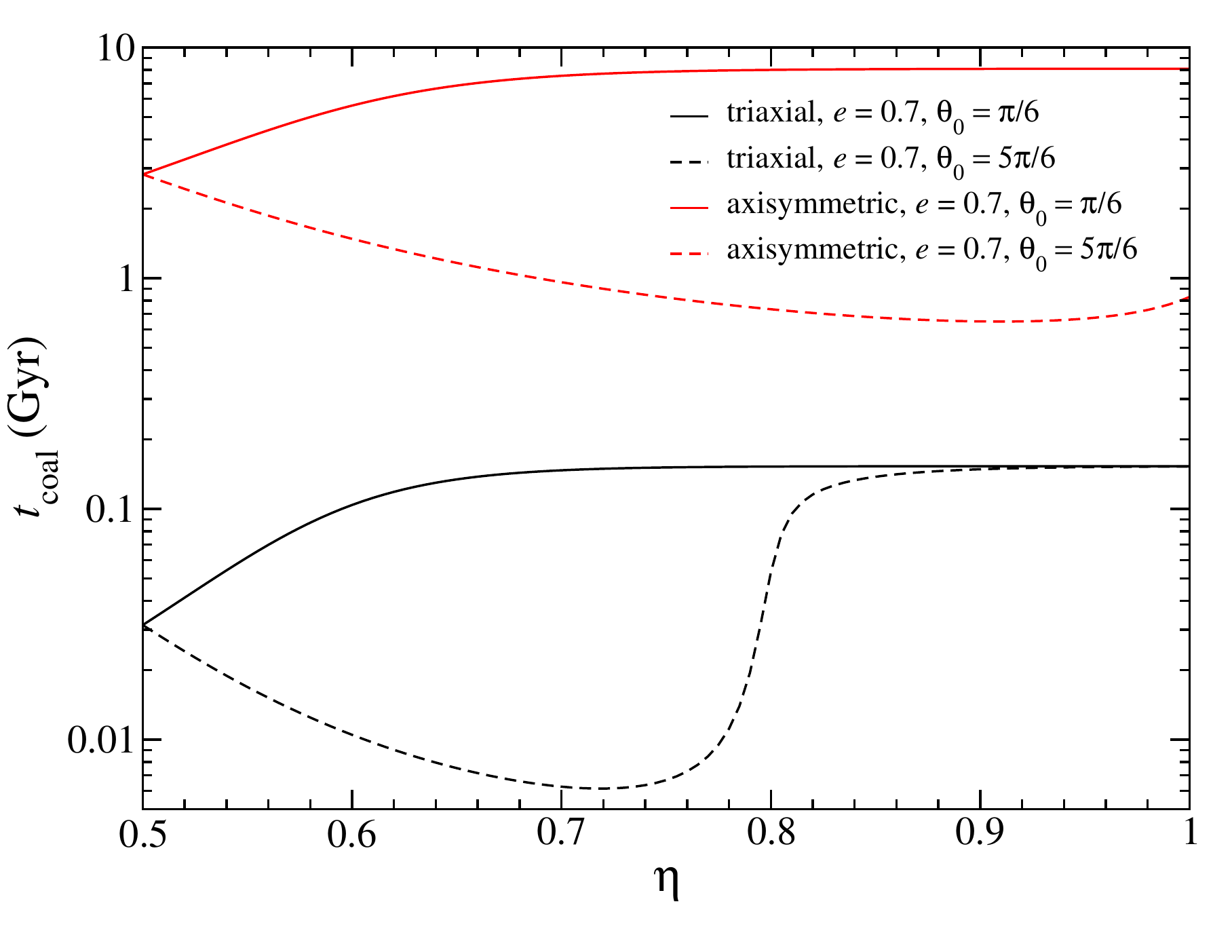}}
\caption{
The dependence of strain amplitude for axisymmetric galaxy model (up) and coalescence time for triaxial and axisymmetric galaxies (down) on the initial conditions. 
}
\label{Figure:A_axi,e>0}
\end{figure}

\begin{figure}[h!]
	\centering
	\subfigure{\includegraphics[width=0.49\textwidth]{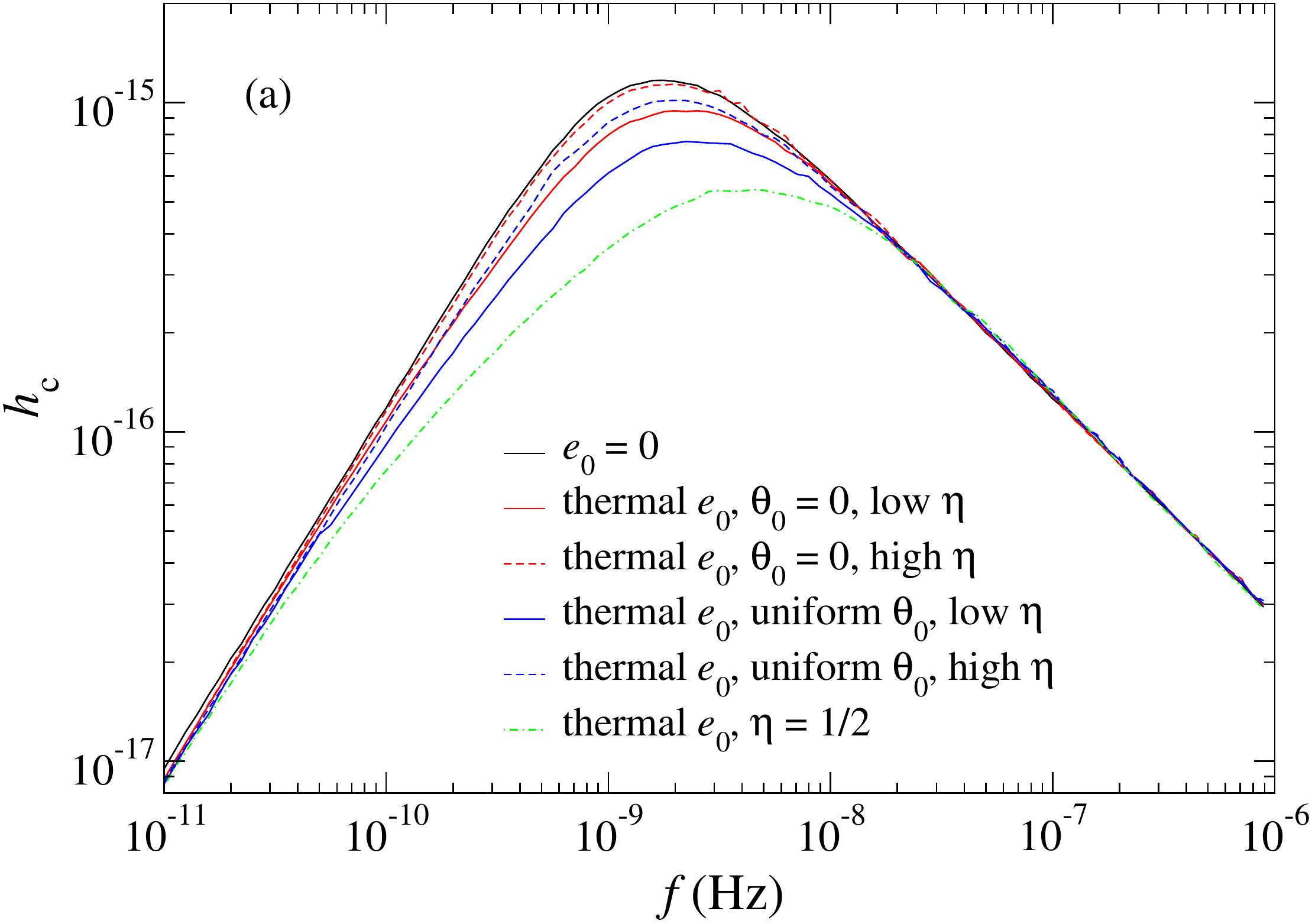}}
	\subfigure{\includegraphics[width=0.49\textwidth]{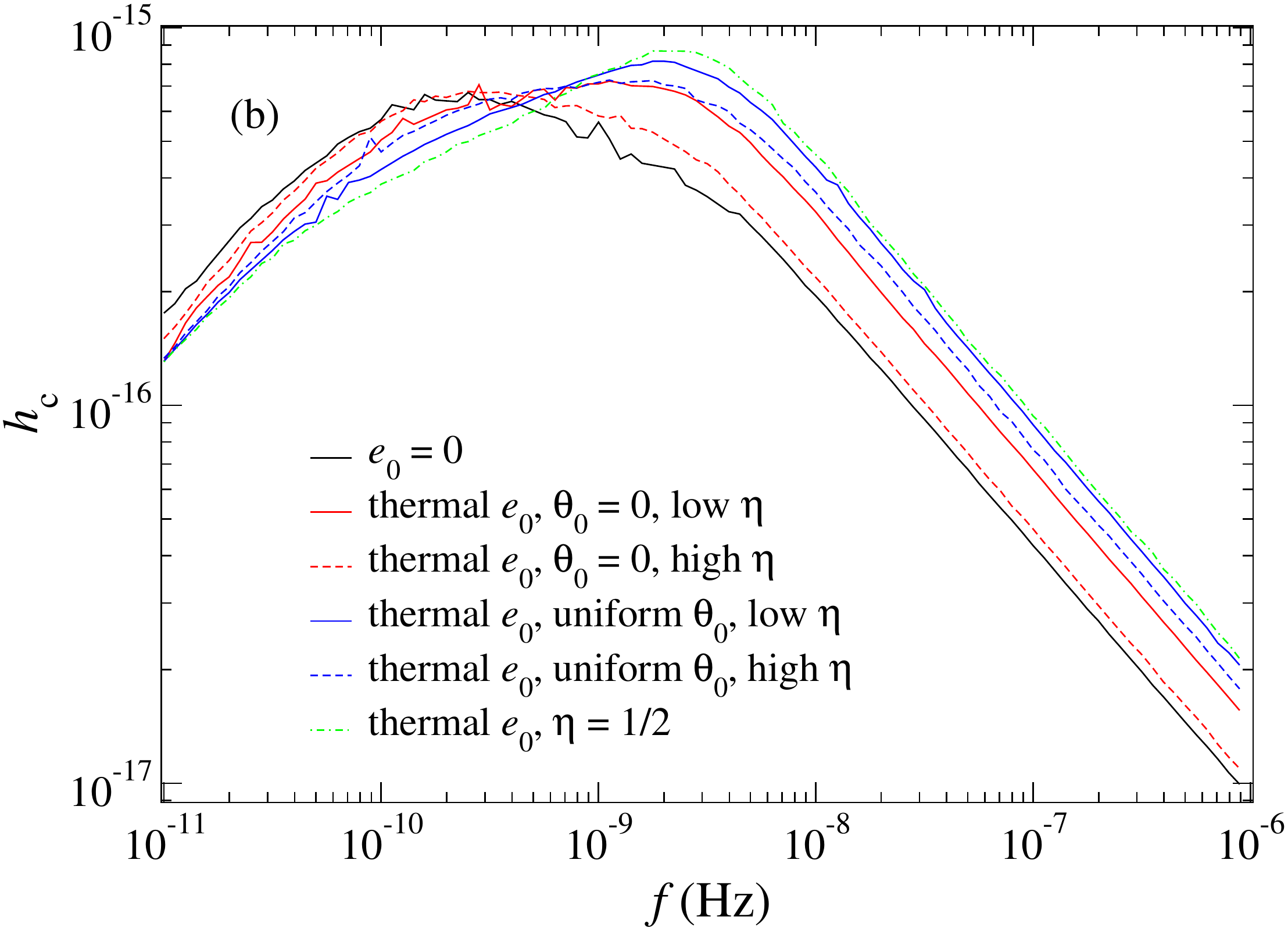}}
\caption{
Predicted GW strain assuming $\beta=10^{-3}$, ``thermal'' distributions of initial eccentricity, 
two different distributions of $\theta_0$ (Eq. \ref{Equation:e_0-thermal}) and two different distributions of $\eta$ 
(``low $\eta$'',  Eq.~(\ref{Equation:vsigmasesana}a); ``high $\eta$'', Eq. (\ref{Equation:vsigmasesana}b))
for (a) triaxial and (b) axisymmetric galaxies. 
Also shown for comparison is the curve computed assuming a thermal eccentricity distribution 
and no rotation ($\eta=1/2$).
}
\label{Figure:h_c-final}
\end{figure}

Finally, we construct a model which assumes that all spiral galaxy bulges are axisymmetric and fast-rotating,
and that elliptical galaxies have different shapes depending on their mass:
galaxies smaller than $10^{11.25}\msun$ (``fast rotators'') are axisymmetric while those heavier than $10^{11.25}\msun$ (``slow rotators'') are triaxial.
This dichotomy is motivated observationally by the different morphologies and shape distributions of galaxies in the two
mass ranges \cite{TremblayMerritt1996,Emsellem2011}. For spirals and fast rotators we assume the $\eta$ distribution from Eq.~(\ref{Equation:vsigmasesana}a), for slow rotators -- the one from Eq.~(\ref{Equation:vsigmasesana}b).

\begin{figure}[h!]
        \centering
        \subfigure{\includegraphics[width=0.49\textwidth]{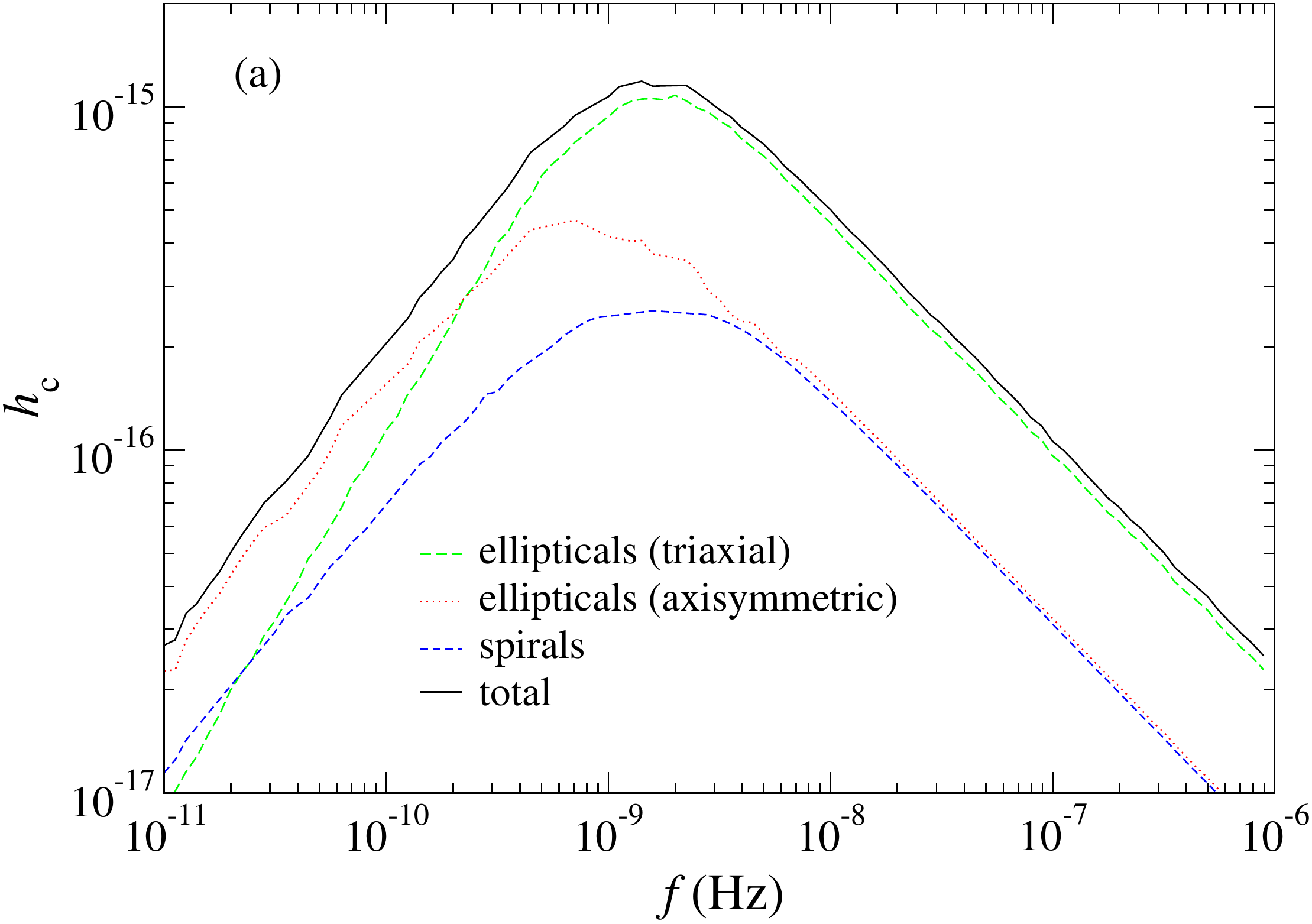}}
	\subfigure{\includegraphics[width=0.49\textwidth]{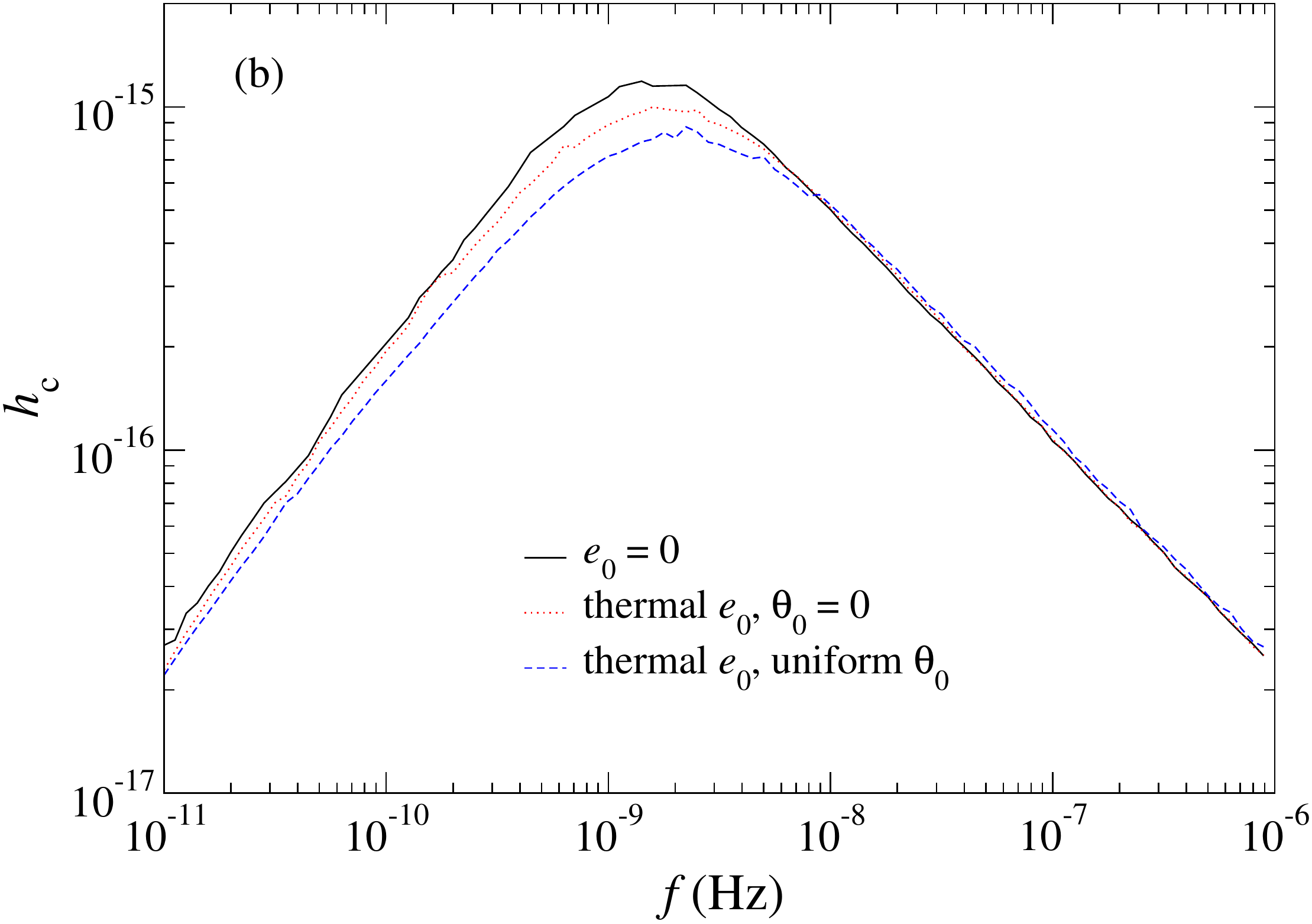}}
\caption{
(a) Contribution of different galaxy types to the predicted GW strain assuming $\beta=10^{-3}$ and zero eccentricity for all binaries. 
(b) Strain spectra including the mixture of galaxy types described in the text, for different assumed distributions of $e_0$ and $\theta_0$.}
\label{Figure:h_c(galaxy_type)}
\end{figure}

Fig.~\ref{Figure:h_c(galaxy_type)} shows the results.
Fig.~\ref{Figure:h_c(galaxy_type)}a plots the contributions from different galaxy types assuming $e=0$.
The signal at PTA frequencies is heavily dominated by triaxial (elliptical) galaxies --
not surprising considering that they are the most massive galaxies. 
Because of that, the dependence of the total signal on the distributions of $e_0$ and $\theta_0$ 
(Fig.~\ref{Figure:h_c(galaxy_type)}b) is almost identical to that for triaxial galaxies (Fig.~\ref{Figure:h_c-final}a), and its amplitude at high frequencies is practically independent of $e_0$, $\theta_0$:
\barr
\ayr &\approx& 2.31\times10^{-16}\, \left(\frac{\beta}{10^{-3}}\right)^{0.85}
\earr

Setting $\beta=10^{-3}$, our preferred value, would reduce the GW strain amplitude by a factor $\sim3$ compared to previous estimates,
enough to account for the discrepancies between the models and the current PTA upper limits.

\section{Discussion}
\label{Section:Discussion}

We have presented calculations of the isotropic gravitational wave (GW) background spectrum that would be produced by a population
of binary supermassive black holes (\sbhs) in galactic nuclei. 
In our model, massive binaries evolve at large separations due to interaction with their stellar environment 
and at small separations due to emission of GWs.
New features of our calculation, and the major results, are summarized here.

\begin{enumerate}

\item We model the time dependence of the binary hardening rate, $S=(d/dt)(1/a)$, using the results of \citet{Vasiliev2014,Vasiliev2015} who derived expressions for $s(a)$ that are valid in the large-$N$
(collisionless) limit appropriate to giant galaxies. 
These expressions imply efficient coalescence for binaries at the centers of triaxial (i.e. non-axisymmetric)
galaxies, like those that are expected to form in galaxy mergers.
In the case of axisymmetric geometries -- which may be a better representation of low-luminosity galaxies -- 
binary hardening rates are predicted to be lower  (Fig.~\ref{Figure:t,A,e=0}a), implying a dependence of 
coalescence timescale $\tcoal$ on galaxy luminosity, hence on $M_\mathrm{BH}$.

\item Rapid evolution of binary \sbhs\ in triaxial galaxies significantly (by a few orders of magnitude) 
decreases GW emission at low frequencies, $f\lesssim10^{-9}\,\mathrm{Hz}$, compared with a fiducial model in which
evolution is driven entirely by GW emission itself.
Evolution timescales are short enough in this geometry ($\tcoal\lesssim300\,\mathrm{Myr}$) that essentially all
binaries would reach coalescence, hence $h_c(f)$ at high frequencies includes contributions from essentially every
binary that forms, and it obeys the standard $h_c\propto f^{-2/3}$ dependence for $f\gtrsim3\cdot10^{-9}\,\mathrm{Hz}\approx0.1\,\mathrm{yr}$, which is approximately the current PTA sensitivity range \citep{Lentati2015,Shannon2015,Arzoumanian2016}).
In axisymmetric galaxies, binary evolution at large separations is slower.
The frequency below which the $h_c\sim f^{-2/3}$ spectrum is truncated is  $f\lesssim10^{-10}\,\mathrm{Hz}$ in this case; 
furthermore, since $\tcoal$ can be longer than a Hubble time for $M_{12}\gtrsim4\times10^8\msun$ 
(the ``final-parsec problem''), there is a reduction in the contribution of these binaries to $h_c(f)$ at high frequencies as well,
lowering the predicted amplitude of $h_c(f)$ (however, the situation is different for highly eccentric binaries which have much lower $\tcoal$; see Fig.~\ref{Figure:A_axi,e>0}).

\item Galactic nuclei are generically rotational in the sense that there is a preferred axis about which stars orbit.
Eccentricity evolution of a massive binary in a rotating nucleus depends strongly on its initial angular momentum 
direction compared with that of the nucleus \citep{PaperI}.
Initially corotating binaries as well as some of the counterrotating ones circularize very quickly due to stellar encounters and enter
the PTA band while almost circular. 
Counterrotating binaries can attain very high eccentricities ($e>0.9$) if their angular momenta are initially strongly inclined with respect to the nucleus and in some cases will retain this high eccentricity even when entering the GW-dominated regime 
(Fig.~\ref{Figure:e>0}).
High eccentricities imply a reduction in $h_c(f)$ at low frequencies;
in the case of axisymmetric galaxies it also increases $h_c(f)$ at high frequencies (by a factor as great as $\sim2$) because it shortens the coalescence time, allowing more binaries to enter GW emission regime and contribute to the PTA signal.

\item We argue (\S~\ref{Section:mbh-mbulge}) that previous calculations of $h_c(f)$ have been based 
on an incorrect (over-estimated) value of  $M_\mathrm{BH}/M_\mathrm{bulge}$, the mean ratio of \sbh\ mass to bulge mass.
We adopt a fiducial value of $0.001$ for this ratio, compared with $\sim 0.003$ in most other studies.
This lower value results in a reduction in the predicted $h_c(f)$ at all frequencies 
(since $L_\mathrm{GW} \propto M_{12}^{10/3}$), and a shift in the peak of $h_c(f)$ to higher frequencies 
(lower-mass \sbhs\ enter the GW-dominated regime at higher orbital frequencies). 
We show that in the frequency regime currently accessible to PTAs, $h_c(f)\propto (M_{12}/M_\mathrm{bulge})^{0.85}$ (Eq. \ref{Equation:A_yr(beta)} and Fig. \ref{Figure:t,A,e=0}b), so that our choice for this ratio implies a factor $\sim 2$ reduction in 
the characteristic strain.
One consequence is that the existing ``tension'' between theoretical predictions of $h_c(f)$ and observational non-detections
by PTAs \citep{Shannon2015,Arzoumanian2016} is removed.
\end{enumerate}

\citet{Shankar2016} argued that a selection bias exists such that almost all galaxies in which the \sbh\ influence sphere has been resolved have velocity dispersions that are higher than average for a fixed galaxy mass. The impact of this bias on detection of GWs by PTAs  was analyzed in \cite{Sesana2016}. Both of these studies accept at face value claims that \sbh\ influence radii have been resolved. We discuss, in \S~\ref{Section:mbh-mbulge}, why this assumption is likely to be incorrect.

For the values of $M_\mathrm{BH}/M_\mathrm{gal}$ that we favor,
almost all of our models predict $A_\mathrm{yr} < 5\times 10^{-16}$. 
This value of $A_\mathrm{yr}$ was identified as a
``conservative {\it lower} limit'' (emphasis added) by \citet{Siemens2013} in their study of 
time-to-detection of the stochastic GW background by PTAs.
Those authors considered two ways in which PTA detection limits might improve over time: due to lengthened
data streams for individual pulsars, and due to the addition of new pulsars.
Assuming an average of three new pulsars per year (which they considered conservative), 
they estimated a probable date of first detection of GWs of $\sim 2021$ for $A_\mathrm{yr} = 5.6\times 10^{-16}$,
and somewhat later if red noise is present in the auto-correlations.

In our assumed \sbh\ mass -- bulge mass relation, and our expressions for the bulge fractions in different galaxy types, we ignored scatter, assuming the relations to be exact. As Figure 9 of \citet{Simon2016} demonstrates, the presence of scatter in the \sbh\ mass -- bulge mass relation does not simply increase scatter in the computed $h_c(f)$ or $\ayr$; it also increases their mean values. 
In this sense, our results could be viewed as a lower limit on $h_c(f)$.

Even if we accept the pessimistic view that detection of the isotropic GW background by PTAs may lie many years in the future,
sufficiently massive or nearby systems may rise above the stochastic background signal and be individually detectable.
Identification of the binary's host galaxy, and detection of electromagnetic radiative processes  associated with the late evolution
of the binary, can assist in the extraction of the binary parameters from PTA data. 
 Photometric or spectroscopic measurement of the host galaxy's cosmological redshift would provide a distance estimate,
allowing a chirp mass to be derived even for a binary \sbh\ that does not evolve
in frequency over PTA observing timescales \citep{Burke2013}.
These ``multi-messenger'' studies may hold the greatest hope for finally establishing the \sbh\ mass-to-host-mass relations.

\acknowledgements

We thank J. Ellis, A. Graham, L. Lentati, J. Schnittman, A. Sesana, R. Shannon, E. Vasiliev and M. Zemcov for helpful comments and advice. This work was supported by the National Science Foundation under grant no. AST 1211602 and by the National Aeronautics and Space Administration under grant no. NNX13AG92G.


\end{document}